%% file: draft.tex
\documentclass[useAMS,usenatbib]{mnras}
\usepackage{mn2e-breakabs}
\usepackage{epsfig}
\usepackage{amsfonts}
\usepackage[usenames,dvipsnames]{xcolor}
\usepackage{url}
\usepackage{subfigure}
\usepackage[english]{babel}
\usepackage{times,graphicx,amsmath,amsfonts,amssymb,aas_macros,epstopdf,hyperref}
\usepackage[normalem]{ulem}
\usepackage[T1]{fontenc}
\usepackage{multirow}
\usepackage{ae,aecompl}
\usepackage{xspace} 
\usepackage{units}
\usepackage{wasysym}
\usepackage{newtxtext,newtxmath}
\usepackage{arydshln}

\definecolor{ultramarine}{rgb}{0.07, 0.04, 0.56}
\definecolor{cadmiumgreen}{rgb}{0.0, 0.42, 0.24}
\definecolor{indigo}{rgb}{0.0, 0.25, 0.42}

\hypersetup{
    colorlinks=true,
    linkcolor=blue,
    citecolor=blue,
    urlcolor=blue,
}

\usepackage{lineno}

\newcommand{\tc}{\color{black}}
\newcommand{\remove}[1]{}

\def\etal{{\frenchspacing\it et al.}}
\def\ie{{\frenchspacing\it i.e.}}

\def\be{\begin{equation}}
\def\ee{\end{equation}}
\def\ba{\begin{eqnarray}}
\def\ea{\end{eqnarray}}

\def\d{{\rm d}}
\def\dln{{\rm d log}}
\def\m{{\rm M}}
\def\rd{r_{\rm d}}
\def\rfd{r^{\rm fid}_{\rm d}}
\def\zp{z_{\rm p}}

\def\W{{\bf W}}

\def\dd{{\delta\delta}}
\def\tt{{\theta\theta}}
\def\dt{{\delta\theta}}
\def\kunit{{h \ {\rm Mpc}^{-1}}}
\def\abot{{\alpha_{\bot}}}
\def\apar{{\alpha_{\|}}}
\def\fs{{f\sigma_8}}
\def\Hf{{H_{\rm f}}}

\def\zeff{{z_{\rm eff}}}

\title[A tomographic BAO \& RSD analysis of the eBOSS DR14 quasar sample]{The clustering of the SDSS-IV extended Baryon Oscillation Spectroscopic Survey DR14 quasar sample: a tomographic measurement of cosmic structure growth and expansion rate based on optimal redshift weights}
\author[Zhao, Wang, Saito \etal]{
\parbox{\textwidth}{
Gong-Bo Zhao$^{1,2,3}$\thanks{Email: \url{gbzhao@nao.cas.cn}}\thanks{A Royal Society Newton Advanced Fellow}, Yuting Wang$^{1}$, Shun Saito$^{4}$, H\'ector Gil-Mar\'{\i}n$^{5,6}$, Will J. Percival$^{3}$, Dandan Wang$^{1,2}$, Chia-Hsun Chuang$^{7,8}$, Rossana Ruggeri$^{3}$, Eva-Maria Mueller$^{3}$, Fangzhou Zhu$^{9}$, Ashley J. Ross$^{10,3}$, Rita Tojeiro$^{11}$, Isabelle P\^aris$^{12}$, Adam D. Myers$^{13}$, Jeremy L. Tinker$^{14}$, Jian Li$^{1,2}$, Etienne Burtin$^{15}$, Pauline Zarrouk$^{15}$, Florian Beutler$^{3,16}$, Falk Baumgarten$^{7,17}$, Julian E. Bautista$^{18}$, Joel R. Brownstein$^{18}$, Kyle S. Dawson$^{18}$, Jiamin Hou$^{19,20}$, Axel de la Macorra$^{21}$,  Graziano Rossi$^{22}$, John A. Peacock$^{23}$, Ariel G. S\'anchez$^{20}$, Arman Shafieloo$^{24,25}$, Donald P. Schneider$^{26,27}$, Cheng Zhao$^{28,1}$} \\
\vspace*{30pt} \\
$^1$ National Astronomy Observatories, Chinese Academy of Sciences, Beijing, 100012, P.R.China\\
$^2$ University of Chinese Academy of Sciences, Beijing 100049,
China \\
$^3$ Institute of Cosmology \& Gravitation, University of Portsmouth, Dennis Sciama Building, Portsmouth, PO1 3FX, UK\\
$^4$ Max-Planck-Institut f\"ur Astrophysik, Karl-Schwarzschild-Str. 1, 85741 Garching, Germany\\
$^5$ Sorbonne Universit\'es, Institut Lagrange de Paris (ILP), 98 bis Boulevard Arago, 75014 Paris, France\\
$^6$ Laboratoire de Physique Nucl\'eaire et de Hautes Energies, Universit\'e Pierre et Marie Curie, 4 Place Jussieu, 75005 Paris, France\\
$^7$ Leibniz-Institut f\"ur Astrophysik Potsdam (AIP), An der Sternwarte 16, D-14482 Potsdam, Germany\\
$^8$ Kavli Institute of Particle Astrophysics and Cosmology \& Physics Department, Stanford University, Stanford, CA 94305, USA\\ 
$^9$ Department of Physics, Yale University, 260 Whitney Ave, New Haven, CT 06520, USA\\
$^{10}$ Center for Cosmology and Astro-Particle Physics, Ohio State University, Columbus, Ohio, USA\\
$^{11}$ School of Physics and Astronomy, University of St Andrews, St Andrews, KY16 9SS, UK\\
$^{12}$ Aix-Marseille Universit\'e, CNRS, LAM (Laboratoire d'Astrophysique de Marseille), 38 rue F. Joliot-Curie 13388 Marseille Cedex 13, France\\
$^{13}$ Department of Physics and Astronomy, University of Wyoming, Laramie, WY 82071, USA\\
$^{14}$ Center for Cosmology and Particle Physics, Department of Physics, New York University, New York, NY 10003, USA\\
$^{15}$ IRFU,CEA, Universit\'e Paris-Saclay, F-91191 Gif-sur-Yvette, France\\
$^{16}$ Lawrence Berkeley National Lab, 1 Cyclotron Rd, Berkeley CA 94720, USA\\
$^{17}$ Humboldt-Universit\"at zu Berlin, Institut f\"ur Physik, Newtonstrasse 15,D-12589, Berlin, Germany\\
$^{18}$ Department Physics and Astronomy, University of Utah, 115 S 1400 E, Salt Lake City, UT 84112, USA\\
$^{19}$ Universit\"{a}ts-Sternwarte M\"{u}nchen, Ludwig-Maximilians-Universit\"{a}t M\"{u}nchen, Scheinerstra{\ss}e 1, 81679 M\"{u}nchen, Germany\\
$^{20}$ Max-Planck-Institut f\"ur Extraterrestrische Physik, Postfach 1312, Giessenbachstr., 85748 Garching bei M\"unchen, Germany\\
$^{21}$ Instituto de F\'{\i}sica, Universidad Nacional Aut\'onoma de M\'exico, Apdo. Postal 20-364, M\'exico\\
$^{22}$ Department of Physics and Astronomy, Sejong University, Seoul 143-747, Korea\\
$^{23}$ Institute for Astronomy, University of Edinburgh, Royal Observatory, Blackford Hill, Edinburgh, UK\\
$^{24}$ Korea Astronomy and Space Science Institute, Yuseong-gu, 776 Daedeok daero, Daejeon 34055, Korea\\
$^{25}$ University of Science and Technology, Yuseong-gu 217 Gajeong-ro, Daejeon 34113, Korea\\
$^{26}$ Department of Astronomy and Astrophysics, The Pennsylvania State University, University Park, PA 16802, USA\\
$^{27}$ Institute for Gravitation and the Cosmos, The Pennsylvania State University, University Park, PA 16802, USA\\
$^{28}$ Tsinghua Center for Astrophysics and Department of Physics, Tsinghua University, Beijing 100084, China
}
\date{Accepted XXX. Received YYY; in original form ZZZ}
\pubyear{2018}

\begin{document}
\label{firstpage}
\pagerange{\pageref{firstpage}--\pageref{lastpage}}
\maketitle

\begin{abstract} 

We develop a new method, which is based on the optimal redshift weighting scheme, to extract the maximal tomographic information of baryonic acoustic oscillations (BAO) and redshift space distortions (RSD) from the extended Baryon Oscillation Spectroscopic Survey (eBOSS) Data Release 14 quasar (DR14Q) survey. We validate our method using the EZ mocks, and apply our pipeline to the eBOSS DR14Q sample in the redshift range of $0.8<z<2.2$. We report a joint measurement of $f\sigma_8$ and two-dimensional BAO parameters $D_{\rm A}$ and $H$ at four effective redshifts of $z_{\rm eff}=0.98, 1.23, 1.52$ and $1.94$, and provide the full data covariance matrix. Using our measurement combined with BOSS DR12, MGS and 6dFGS BAO measurements, we find that the existence of dark energy is supported by observations at a $7.4\sigma$ significance level. Combining our measurement with BOSS DR12 and Planck observations, we constrain the gravitational growth index to be $\gamma=0.580\pm0.082$, which is fully consistent with the prediction of general relativity. This paper is part of a set that analyses the eBOSS DR14 quasar sample.
\end{abstract}


\section{Introduction}

In this era of precision cosmology, large spectroscopic galaxy surveys are one of the key probes of both the expansion history and structure growth of the Universe. Probing deep into the Universe, these surveys are able to provide rich information on the past lightcone, which is crucial to unveil the physics of the cosmic acceleration \citep{Riess,Perlmutter}, through studies of dark energy \citep{DEreview} and gravity on cosmological scales \citep{MGreview}.

{\tc Baryonic acoustic oscillations (BAO) and redshift space distortions (RSD) are distinct three-dimensional clustering patterns probed by galaxy surveys, which are key to map the expansion history and the structure growth of the Universe respectively. Since a first successful measurement of BAO in 2005 \citep{Eisenstein05} and RSD in 2001 \citep{2dfrsd}, measurements with higher precision have been performing actively using large galaxy surveys \citep{sdss2bao,6dF,6dfrsd,wigglezrsd,wigglezbao,mgs,desy1,alam,dr14lrg,DR14BAO}.}

Traditional BAO and RSD measurements are usually performed in a single, or a small number of redshifts slices, which is to guarantee that there are sufficiently large number of galaxies for the analysis to avoid large statistical or systematic uncertainties. However, this approach may give rise to information loss of the temporal evolution of the BAO or RSD signal, which is essential for tests of cosmological models. One solution to this problem is to perform BAO and RSD analysis in a large number of overlapping redshift slices to balance the level of uncertainty for the BAO/RSD analysis and the tomographic information \citep{Zhaotomo16, Wangtomo17, Wangtomo16}. However, this method is computationally expensive as it requires repetitive measurements and analysis with the computational cost scaling with $N_{z}(N_z-1)/2$, where $N_z$ is the number of redshift slices.

The optimal redshift weighting scheme, which was first developed for cosmological implications by \cite{Fisher}, is a computationally efficient alternative. By designing the optimal redshift weights for a given set of parameters, one can in principle extract the lightcone information by fewer than $N_p$ measurements, where $N_p$ is the number of parameters to be measured. Given that $N_p$ is usually a small number for BAO and RSD analysis, this approach significantly reduces the computational cost. 

The optimal redshift method has been applied to BAO measurements in configuration space \citep{ZPW, zwBAOmock, Zhu18}, and RSD measurements in Fourier space \citep{zwRSD, RR18}. In this work, we develop an alternative approach to \cite{zwRSD, RR18} for a joint measurement of BAO and RSD in Fourier space, and apply our method to the extended Baryon Oscillation Spectroscopic Survey (eBOSS) Data Release 14 quasar (DR14Q) sample, followed by a cosmological implication.

The paper is structured as follows. In Section \ref{sec:data}, we describe the observational and simulated datasets used in this analysis, and in Section \ref{sec:method}, we present the method, followed by mock tests and main result of this work in Section \ref{sec:result}. We compare our BAO and RSD measurement to the DR14Q companion papers presented in Section \ref{sec:consensus}, followed by a cosmological implication of our measurement in Section \ref{sec:cosmology}, before conclusion and discussion in Section \ref{sec:conclusion}.

\section{The datasets}
\label{sec:data}

In this section, we briefly describe the observational and simulated datasets used in this analysis. We refer the readers to a more detailed description of the DR14Q datasets in a companion paper of \cite{HGM18}.

\subsection{The eBOSS DR14Q sample} 

Being part of the Sloan Digital Sky Survey-IV (SDSS-IV) project \citep{sdss4tech}, the eBOSS quasar survey \citep{ebossoverview, eBOSSZhao16} started in 2014 using a 2.5-metre Sloan telescope \citep{sdsstelescope} at the Apache Point Observatory in New Mexico in the United States. After the eBOSS quasar target selection, which is described in \cite{ebossqsoselection}, the spectra are taken using the double-armed spectrographs \citep{sdssspectrograph}, which were used for the Baryon Oscillation Spectroscopic Survey (BOSS) mission, as part of the SDSS-III project \citep{sdss3tech}.

\begin{figure}
\centering
{\includegraphics[scale=0.3]{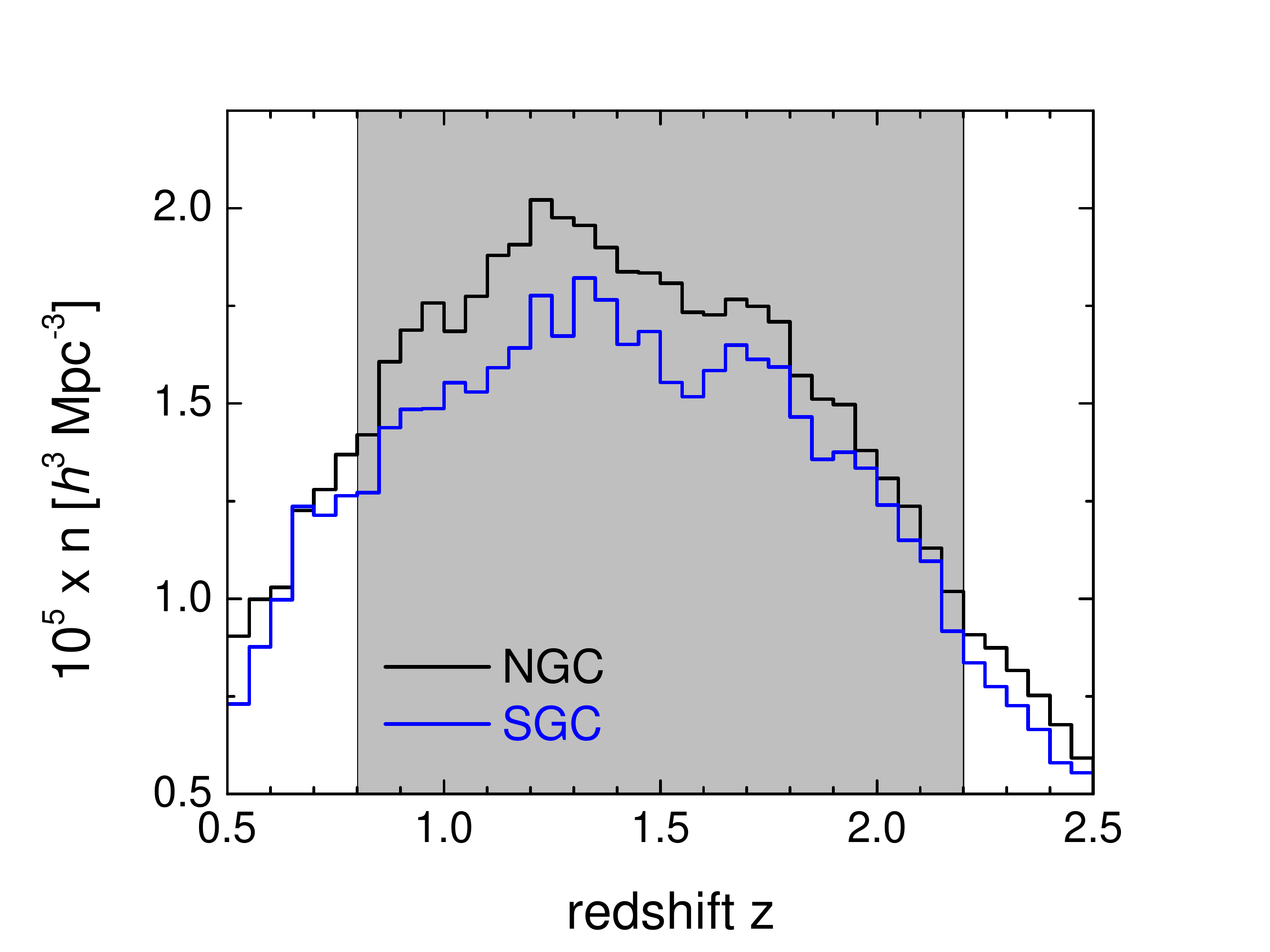}}
\caption{The observed volume number density (binned with $\Delta z=0.05$) of the quasars in unit of $h^3 \ {\rm Mpc}^{-3}$ (multiplied by $10^5$) as a function of redshifts in the NGC (upper black) and SGC (lower blue). The grey shaded region ($0.8<z<2.2$) shows the redshift range in which data are selected for this analysis.}
\label{fig:nz}
\end{figure}

The data catalogue used in this analysis is the eBOSS quasar sample \citep{QSOcat}, which is a part of the SDSS-IV Data Release 14 \citep{dr14}. This DR14Q catalogue consists of around 150, 000 quasars with secure redshifts distributed across an effective area of 2112.9 deg$^2$ ({\tc see Figure 3 in a companion paper \citealt{HGM18} for a footprint of the DR14Q sample}). A histogram for the redshift distribution for the quasar sample is shown in Fig. \ref{fig:nz}. Each quasar is given a total weight of \ba \label{eq:wtot} w_{\rm tot} = w_{\rm FKP} \ w_{\rm sys} \ w_{\rm spec}  \sqrt{w_z} \ . \ea where $w_{\rm FKP}, w_{\rm sys}$ and $w_{\rm spec}$ denotes for the Feldman-Kaiser-Peacock (FKP) weight \citep{FKP}, systematics weight and the spectrum weight. The FKP weight is used to minimise the uncertainty of the power spectrum measurement, and $w_{\rm sys}$ corrects for the systematic effects from observing conditions including seeing, airmass, extinction, sky background and so on \citep{DR14BAO}. The spectrum weight accounts for the fibre collision and redshift failures \citep{HGM18,PZ18}. In addition, we apply a redshift weight to each quasar to capture the tomographic information in redshift, which is detailed in Sec. \ref{sec:zweight} \footnote{As the redshift weights derived in Sec. \ref{sec:zweight} are for power spectrum multipoles, each quasar should be assigned a square root of the weights.}. {\tc The DR14Q sample used in this analysis is publicly available on the SDSS website \url{https://data.sdss.org/sas/dr14/eboss/lss/catalogs/}}

\subsection{The simulated mock samples} 

A large number of mock samples, each of which has the same clustering property of the eBOSS DR14Q sample, are required to estimate the data covariance matrix. In this analysis, we use the Extended Zel'dovich (EZ) mocks, which consist of $1000$ realisations, produced following the prescription in \cite{EZmock}. The cosmological parameters used for the EZ mocks are listed in Eq (\ref{eq:pez}), where the parameters are: the physical energy density of cold dark matter and baryons, the sum of neutrino masses, the amplitude of the linear matter power spectrum within $8 h^{-1} \ {\rm Mpc}$, the power index of the primordial power spectrum, and the (derived) scale of the sound horizon at recombination respectively.

\ba {\bf \Theta}&\equiv&\left\{\Omega_c h^2, \Omega_b h^2, \sum M_{\nu}/{\rm eV}, \sigma_8, n_s, \rd/{\rm Mpc} \right\}   \nonumber \\
\label{eq:pez}&= &\{0.1189,0.0221,0,0.8225,0.96,147.66\}|_{\rm EZ}   \\
\label{eq:pfid}&= &\{0.1190,0.022,0.06,0.8,0.97,147.78\}|_{\rm f}  \ea

We list another set of parameters in Eq (\ref{eq:pfid}), which is the fiducial cosmology we adopt for this analysis \footnote{Throughout the paper, the subscript or superscript `${\rm f}$' denotes the fiducial value.}. 

{\tc Note that the EZ mocks used in this analysis include the full information on the lightcone, which is essential for the tomographic analysis in this study. The lightcone-mocks were constructed by stacking simulation boxes at various redshifts. In order to match the time-evolution of the clustering signal of the DR14Q sample, parameters used for the mocks were calibrated from the DR14Q sample in overlapping redshift slices, which is necessary to reduce the noise. For more details of the production of lightcone mocks, please refer to Section 5.1 of \cite{DR14BAO}.} 

\section{Methodology}
\label{sec:method}

In this section, we present details of the method used for this analysis, including the parametrisation, the derivation of the optimal redshift weights, the template, and the likelihood analysis with details on parameter estimation. 

We start by parametrising the lightcone information of redshift surveys using a small number of parameters, and aim to derive a set of redshift weights to optimise the measurement of these parameters simultaneously. 

\subsection{Parametrising tomographic information in redshift surveys}

\subsubsection{Parametrisation of the BAO parameters}

As in \cite{ZPW,zwBAOmock}, we parametrise the redshift-dependence of the transverse and radial dilation of the BAO distances $\alpha_{\bot}$ and $\alpha_{\|}$ using the following form,
\ba\label{eq:alpha} &&\alpha_{\bot}(z)\equiv\frac{D_A(z)}{D_A^{\rm f}(z)}\theta = \alpha_0 \left(1+\alpha_1 x  \right), \nonumber \\ 
&&\alpha_{\|}(z)\equiv\frac{H_{\rm f}(z)}{H(z)} \theta =\alpha_0\left( 1+\alpha_1+2\alpha_1 x  \right), \nonumber \\ 
&& x\equiv\chi_{\rm f}(z)/\chi_{\rm f}(z_{\rm p})-1, \ \theta\equiv{\rd^{\rm f}}/{\rd},\ea where $\chi$ is the comoving distance, and $\alpha_0$ and $\alpha_1$ are free parameters. {\tc This parametrisation is essentially a Taylor expansion, and as stated in \citet{ZPW}, it can well approximate the background expansion history of a wide range of cosmologies.} The pivot redshift $\zp$ is taken to be the effective redshift of the DR14Q sample, which is defined as follows \citep{S14}, 

\ba\label{eq:zeff} \zp = z_{\rm eff} = \frac{\sum_i z_i w_i^2} {\sum_i w_i^2}.   \ea Here $w_{i}$ is the total weight of the $i$th data sample shown in Eq (\ref{eq:wtot}).  
{\tc Note that in Eq (\ref{eq:alpha}), $x$ vanishes at $z=\zp$, which relates $\alpha_0$ and $\alpha_1$ to $\alpha_{\bot}(\zp)$ and $\alpha_{\|}(\zp)$ via,
\ba\label{eq:alpha12} &&\alpha_{\bot}(\zp) = \alpha_0,  \nonumber \\ 
&&\alpha_{\|}(\zp) =\alpha_0\left( 1+\alpha_1\right).  \ea Plugging Eq (\ref{eq:alpha12}) into Eq (\ref{eq:alpha}), one obtains,}
\ba
\label{eq:alpha_perp_z}&&\alpha_{\bot}(z)= \alpha_{\bot}(\zp)+\left[ \alpha_{\|}(\zp) - \alpha_{\bot}(\zp)  \right]x, \nonumber \\
\label{eq:alpha_para_z}&&\alpha_{\|}(z)= \alpha_{\|}(\zp)+2\left[ \alpha_{\|}(\zp) - \alpha_{\bot}(\zp)  \right]x.
\ea

\subsubsection{Parametrisation of the RSD parameters}

We assume that the logarithmic growth rate $f$ takes the form of \citep{Lindergamma},
\ba\label{eq:f} f(z)\equiv\frac{\dln\delta}{\dln a}=\left[\Omega_\m(z)\right]^{\gamma} =\left[\Omega_\m H_0^2(1+z)^3\right]^\gamma\left[\frac{\alpha_{\|}(z)}{H_{\rm f}(z)\theta}\right]^{2\gamma}\ea where $\delta$ is the overdensity.
With $f(z)|_{z=\zp}=f(\zp)$, the above equation can be recast into,
\ba f(z) = f(\zp)\left[\frac{\Omega_\m(z)}{\Omega_\m(\zp)}\right]^{\gamma} \ea where
\ba && \frac{\Omega_\m(z)}{\Omega_\m(\zp)} = \left(\frac{1+z}{1+\zp} \right)^3 \left[\frac{\apar(z)}{\apar(\zp)} \right]^2 \left[ \frac{\Hf(\zp)}{\Hf(z)} \right]^2 \\
&& \Hf(z) \propto \left[\Omega^{\rm f}_\m(1+z)^3+(1-\Omega^{\rm f}_\m)\right]^{1/2}\ea The gravitational growth index $\gamma$ is treated as a free parameter.

The time evolution of the normalisation $\sigma_8$ is modelled as,
\ba\sigma_8(z)=\sigma_8(z_{\rm p})\frac{D(z)}{D(z_{\rm p})}\ea where 
\ba D(z)={\rm exp}\left[-\int_0^z {\rm d}z\frac{f(z)}{1+z}\right]\ea In this framework, the entire evolution history of $f\sigma_8$ is known given parameters $f(\zp), \sigma_8(\zp), \apar(\zp), \abot(\zp)$ and $\gamma$.

\subsubsection{Parametrisation of the bias parameters}

The redshift evolution of the linear bias $b_1$ for the DR14Q sample has been found to be well approximated by a quadratic function \citep{QSObias} developed in \citet{Croom}. In this work, we adopt the fitting formula developed in \cite{Croom} with one parameter $b_1(\zp)$ to be determined, \ie,
\ba b_1(z)=b_1(\zp)+0.29\left[(1+z)^2-(1+z_{\rm p})^2\right]\ea

The time evolution of the nonlocal bias $b_2$ has not been well studied in the literature. As it is expected to be much less important compared to the linear bias on scales of interest for BAO and RSD, we assume it to be a constant for simplicity, \ie, 
\ba b_2(z)=b_2(\zp)\ea

\subsubsection{Parametrisation of the FoG parameter}

The RSD signal is affected by the so-called Finger-of-God (FoG) radial
smearing owing to virialised peculiar velocities. We assume that
the corresponding velocity dispersion, $\sigma_v$, is proportional to $(1+z)/H(z)$ during evolution \citep{SE07}, thus, 
\ba\sigma_v(z) = \frac{\Hf(\zp)}{H_{\rm f}(z)} \frac{\alpha_{\|}(z)}  {\alpha_{\|}(\zp)}  \frac{1+z}{1+\zp} \sigma_v(\zp)\ea

\subsubsection{Summary of the parameters}
\label{sec:param}

The free parameters used with the assumed form of redshift evolution are summerised in Table \ref{tab:param}. In addition to the eight parameters shown in the bottom part of the table, we allocate another parameter $N_{\rm shot}$ to account for the stochasticity of the shot noise of the monopole, \ie, $P_0(k) \rightarrow P_0(k)+N_{\rm shot}$ (see Sections \ref{sec:temp} and \ref{sec:AP} for the definition of $P_0$). 

\begin{table}
\caption{The functional form of the redshift evolution of BAO, RSD and bias parameters used in this work, and their priors. A weak Gaussian prior, which corresponds to the $3 \ \sigma$ constraint derived from Planck 2015 observations \citep{Planck15}, is applied on $\sigma_8(\zp)$.}
\begin{center}
\begin{tabular}{cc}
\hline\hline
Quantities              & redshift evolution   \\
\hline \\
BAO 	& $\alpha_{\bot}(z)= \tc{\alpha_{\bot}(\zp)}+\left[ \tc{\alpha_{\|}(\zp)} - \tc{\alpha_{\bot}(\zp)}  \right]x$  \\
BAO 	& $\alpha_{\|}(z)= \tc{\alpha_{\|}(\zp)}+2\left[ \tc{\alpha_{\|}(\zp)} - \tc{\alpha_{\bot}(\zp)}  \right]x$  \\
\hline
RSD                     &$f(z)=\tc{f(\zp) }\left( \frac{1+z}{1+\zp}  \right)^{3\tc{\gamma}} \left[\frac{\alpha_{\|}(z)}  {\tc{\apar(\zp)}}   \frac{H_{\rm f}(\zp)} {H_{\rm f}(z)}\right]^{2\tc{\gamma}}$   \\
RSD         & $\sigma_8(z)=\tc{\sigma_8(\zp)}\frac{D(z)}{D(z_{\rm p})}$  \\
RSD         & $\sigma_v(z) = \frac{\Hf(\zp)} {\Hf(z)}  \frac{\alpha_{\|}(z)} {\tc{\alpha_{\|}(\zp)} }\frac{1+z}{1+\zp} \tc{\sigma_v(\zp)}$   \\
\hline
Bias   & $b_1(z)=\tc{b_1(\zp)}+0.29\left[(1+z)^2-(1+z_{\rm p})^2\right]$   \\ 
Bias   & $b_2(z) = \tc{b_2(\zp)}$   \\
\hline
Parameter &  Prior \\
\hline
$\tc{\alpha_{\bot}(\zp)}$ & $[0.7,1.3]$ \\
$\tc{\alpha_{\|}(\zp)}$ & $[0.7,1.3]$ \\
$\tc{f({\zp})\sigma_8({\zp})}$ & $[0,  2]$ \\
$\tc{\gamma}$  & $[0, 2]$ \\
$\tc{b_1({\zp})\sigma_8({\zp})}$ & $[0,3]$ \\
$\tc{b_2({\zp})\sigma_8({\zp})}$ & $[-2,2]$ \\
$\tc{\sigma_8(\zp)}$ & $\mathcal{N}(0.367, 0.02^2)$\\
$\tc{\sigma_v(\zp)}$ & $[0,20]$\\
$\tc{N_{\rm shot}}$ & $[-60000, 60000]$\\
\hline\hline

\end{tabular}
\end{center}
\label{tab:param}
\end{table}%

\subsection{The Karhunen-Lo\`eve compression}

\begin{figure}
\centering
{\includegraphics[scale=0.3]{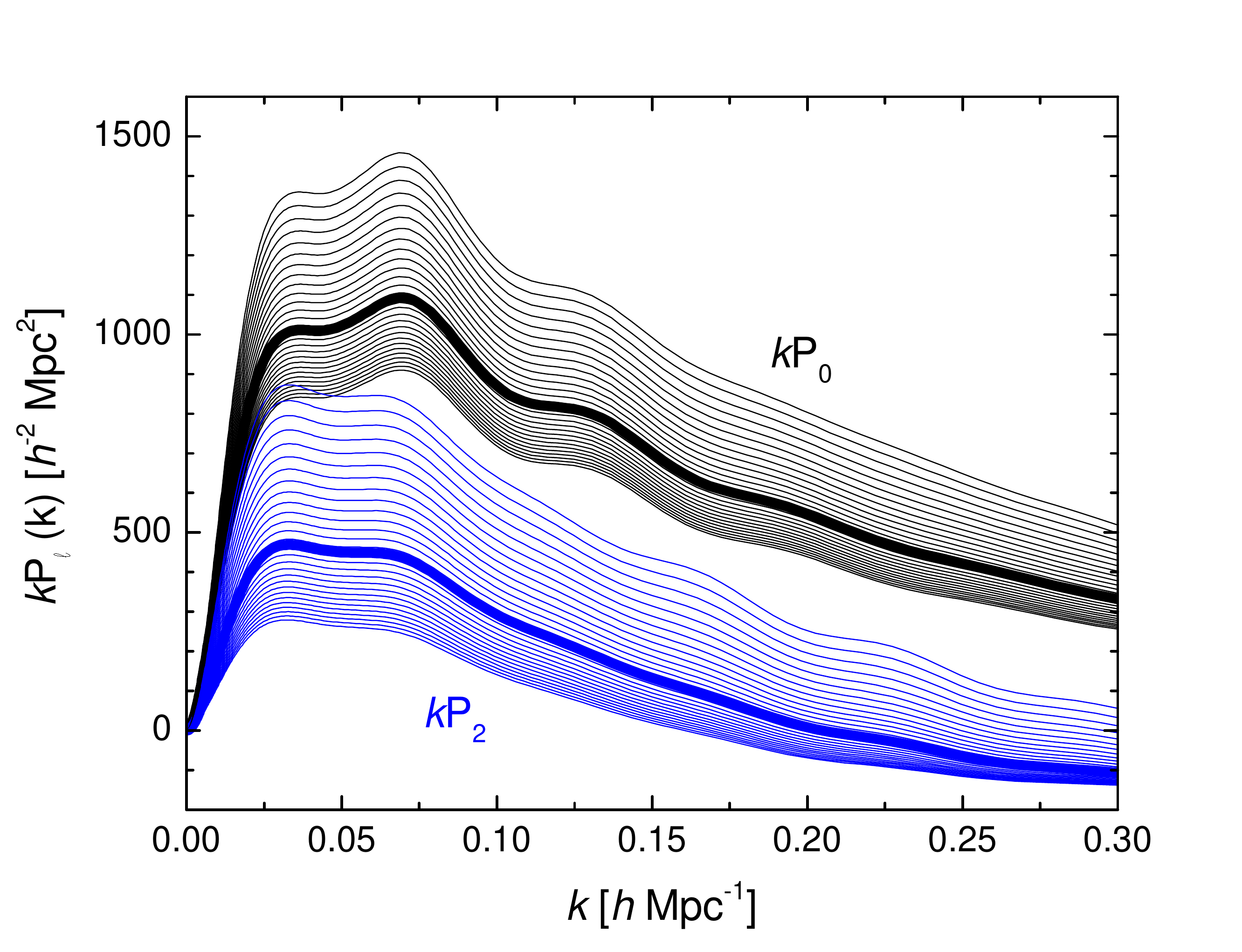}}
\caption{The power spectrum monopole (upper group of curves) and quadrupole (lower group) of the fiducial model at various redshifts. In each group, curves from top to bottom show the fiducial models at redshifts $z=0.825$ to $z=2.175$, with redshift increment of $\Delta z=0.05$. The thick curves within each group show the power spectra of the fiducial model at an effective redshift  of $z_{\rm eff}=1.52$ of the DR14Q sample covering the redshift range of $0.8<z<2.2$.}
\label{fig:pkz}
\end{figure}

\remove{For any given $(k,\mu)$ mode, the Fisher matrix Eq (\ref{eq:Fkmu}) quantifies all the information for the $i$ and $j$th parameters using all the galaxies distributed in multiple redshift slices covering the entire redshift range of $z\in[z_{\rm min}, z_{\rm max}]$.} To analyse the observational data of galaxy surveys, it is impractical to subdivide the galaxies into a large number of redshift slices and perform the measurement in each slice, therefore we seek a way to compress the data sample in redshift with minimum loss of information. 

Data compression by applying optimal redshift weights was recently developed for the BAO measurement \citep{ZPW,zwBAOmock}, based on the Karhunen-Lo\`eve (K-L) compression method \citep{Fisher,Heavens2000}. Here we extend the analysis for a joint measurement of BAO and RSD for redshift surveys.  

To be as general as possible, let us assume that we use $N_p$ parameters to parameterise the galaxy power spectra multipoles in redshift space, which can be in principle measured at $N_z$ redshifts and at $N_k$ wavenumbers. We define the power spectrum vector ${\bf P}$ as, 
\ba  {\bf P}_{\ell,z}(z_i)& \equiv&\left[P_{\ell}(k_1,z_i),P_{\ell}(k_2,z_i),...,P_{\ell}(k_{N_k},z_i)\right]^T\\
       {\bf P}_{z}(z_i)& \equiv&\left[{\bf P}_{0,z}(z_i),{\bf P}_{2,z}(z_i),...,{\bf P}_{2N_{\ell},z}(z_i)\right]^T \\
       {\bf P} &\equiv& \left[{\bf P}_{z}(z_1),{\bf P}_{z}(z_2), ..., {\bf P}_{z}(z_{N_z})\right]^T 
 \ea 
The Fisher information matrix ${\bf F}$ using observables ${\bf P}$ is then, \be {\bf F}={\bf D}^T {\bf C}^{-1}  {\bf D} \ee where ${\bf C}$ is the data covariance matrix, and the derivative matrix ${\bf D}$ is, 
\be\label{eq:D} {\bf D}\equiv\left( \frac{\partial {\bf P}}{\partial p_1}, \frac{\partial {\bf P}}{\partial p_2}, ..., \frac{\partial {\bf P}}{\partial p_{N_p}}\right) \ee

We are seeking an optimal redshift-weighting matrix ${\bf W}$ so that the $z$-weighted power spectra contain the same information for all the parameters. 

The weighted power spectrum vector ($N_pN_kN_{\ell}N_z\times1$) is \be {\bf P}_{\rm w} = {\W}^T {\bf P} \ee where ${\bf W}$ is a $N_zN_{\ell}N_k\times N_p$ weighting matrix, namely, 
\ba \W= 
\left(
\begin{array}{ccc}
W_{0, p_1}(k_1, z_1)    &  ... & W_{0,p_{N_p}}(k_1,z_1)  \\
... &   & ...\\
W_{N_{\ell},p_1}(k_{N_k},{z_{N_z}})    &  ... & W_{N_{\ell},p_{N_p}}(k_{N_k},{z_{N_z}}) \\
\end{array}
\right)
\ea
The data covariance matrix ${\bf C}_{\rm w}$ for the weighted observables ${\bf P}_{\rm w}$ is a ${N_p}\times {N_p}$ matrix, namely, \be {\bf C}_{\rm w}={\bf W}^T~{\bf C}~{\bf W} \ee The Fisher matrix is then, \be\label{eq:Fcomp} {\bf F}_{\rm w}= {\bf D}_{\rm w}^T {\bf C}_{\rm w}^{-1}  {\bf D}_{\rm w}\ee where \be {\bf D}_{\rm w}=\left( \frac{\partial {\bf P}_{\rm w}}{\partial p_1}, \frac{\partial {\bf P}_{\rm w}}{\partial p_2}, ..., \frac{\partial {\bf P}_{\rm w}}{\partial p_{N_p}}\right) ={\bf W}^T {\bf D} \ee The compression is lossless, \ie, ${\bf F}_{\rm w}={\bf F}$, {\tc which means that the information content of a sufficiently redshift-sliced galaxy sample can be made exactly the same as that included in a set of redshift-weighted samples if the redshift weight $\bf W$ is,} \be\label{eq:opW} {\bf W}={\bf C}^{-1} {\bf D}. \ee {\tc In this case, it can be proved that}, \be  {\bf D}_{\rm w}^T ={\bf C}_{\rm w}=  {\bf D}_{\rm w}={\bf F}_{\rm w}= {\bf D}^T {\bf C}^{-1}  {\bf D}={\bf F}. \ee

This is easy to understand qualitatively: to avoid information loss in redshift when measuring $N_p$ parameters at the same time, we have to perform the redshift weighting $N_p$ times using the optimal weight for individual parameters respectively, and use these $N_p$ observables coherently in the likelihood analysis by including the covariance among these observables properly. Note that $N_p=1$, \ie, there is only one parameter to be determined, is a special case where ${\bf P}_{\rm w}, {\bf D}_{\rm w}, {\bf C}_{\rm w}$ and ${\bf F}_{\rm w}$ become scalars, which is the case studied in \cite{ZPW,zwBAOmock,zwRSD}.  

\subsection{The template of power spectrum at a specific redshift}
\label{sec:temp}

We use the extended TNS model \citep{TNS} used in \cite{FBRSD14,FBRSD16,alam} as a template of the two-dimensional power spectrum at a given redshift $z$,

\ba
\label{eq:Pgkmu}  P_{\rm g}(k,\mu,z) &=& {D_{\rm FoG}}\left(k,\mu,z\right)  \left[P_{\rm g,\dd} (k,z)\right. \nonumber \\
                               &&+  2f(z)\mu^2P_{\rm g,\dt} (k,z)  + f^2(z)\mu^4 P_{\rm\tt}(k,z)  \nonumber \\
                               && \left.+ A(k,\mu,z)+B_{}^{}(k,\mu,z)\right],\ea
where 
\ba
P_{\rm g,\dd}(k,z) &=& b_1^2(z) P_{\dd}(k,z)+2b_1(z) b_2(z) P_{\rm b2,\delta}(k,z)   \nonumber \\ 
&&- \frac{8}{7}b_1(z)(b_1(z)-1)P_{\rm bs2,\delta}(k,z)   \nonumber \\
&&+ \frac{64}{315}b_1(z)(b_1(z)-1)\sigma_3^2(k,z)P_m^{\rm L}(k,z)  \nonumber \\
&& + b_2^2(z)P_{\rm b22}(k,z)-\frac{8}{7}[b_1(z)-1]b_2(z)P_{\rm b2s2}(k,z) \nonumber \\
&& + \frac{16}{49}[b_1(z)-1]^2P_{\rm bs22}(k,z),\\
P_{\rm g,\dt}(k,z) &=&b_1(z)P_{\dt}(k,z)+b_2(z)P_{\rm b2,\theta}(k,z)\nonumber \\
&&-\frac{4}{7}[b_1(z)-1]P_{\rm bs2,\theta}(k,z) \nonumber \\
&& + \frac{32}{315}[b_1(z)-1]\sigma_3^2(k,z)P_m^{\rm L}(k,z),\\
P_{\rm g,\tt}(k,z) &=&P_{\tt}(k,z),\\
D_{\rm FoG} (k,\mu,z)&=& \left\{1+\left[k\mu\sigma_v(z)\right]^2/2\right\}^{-2},\\
A(k,\mu,z)&=&b_1^3(z)  \sum_{m,n=1}^3 \mu^{2m} [f(z)/b_1(z)]^n A_{mn}(k,z),  \nonumber \\
B(k,\mu,z)  &=&  b_1^4(z) \sum_{m=1}^4 \sum_{a,b=1}^2 \mu^{2m}[-f(z)/b_1(z)]^{a+b}  B_{ab}^m(k,z). \nonumber\\
\ea

{\tc Note that subscripts $\delta$ and $\theta$ denote the overdensity and velocity divergence fields respectively, and $P_{\dd}, P_{\dt}$ and $P_{\tt}$ are the corresponding nonlinear auto- or cross-power spectrum, which are evaluated using the regularised perturbation theory (RegPT) up to second order \citep{RegPT}. The linear matter power spectrum $P_m^{\rm L}$ is calculated using CAMB \citep{CAMB}. The terms $b_1$ and $b_2$ stand for the linear bias and the second-order local bias respectively. We have eliminated the second-order non-local bias $b_{\rm s2}$ and the third-order non-local bias $b_{\rm 3nl}$ using the following relation \citep{nonlocalb1,nonlocalb2,nonlocalb3},
\ba 
&&b_{\rm s2} = -\frac{4}{7}\left(b_1-1\right),\nonumber \\
&&b_{\rm 3nl} = \frac{32}{315}\left(b_1-1\right).
\ea
The $A$ and $B$ correction terms are computed using standard perturbation theory (SPT) following equations (A3) and (A4) in \citet{TNS}.}


\begin{figure}
\centering
{\includegraphics[scale=0.3]{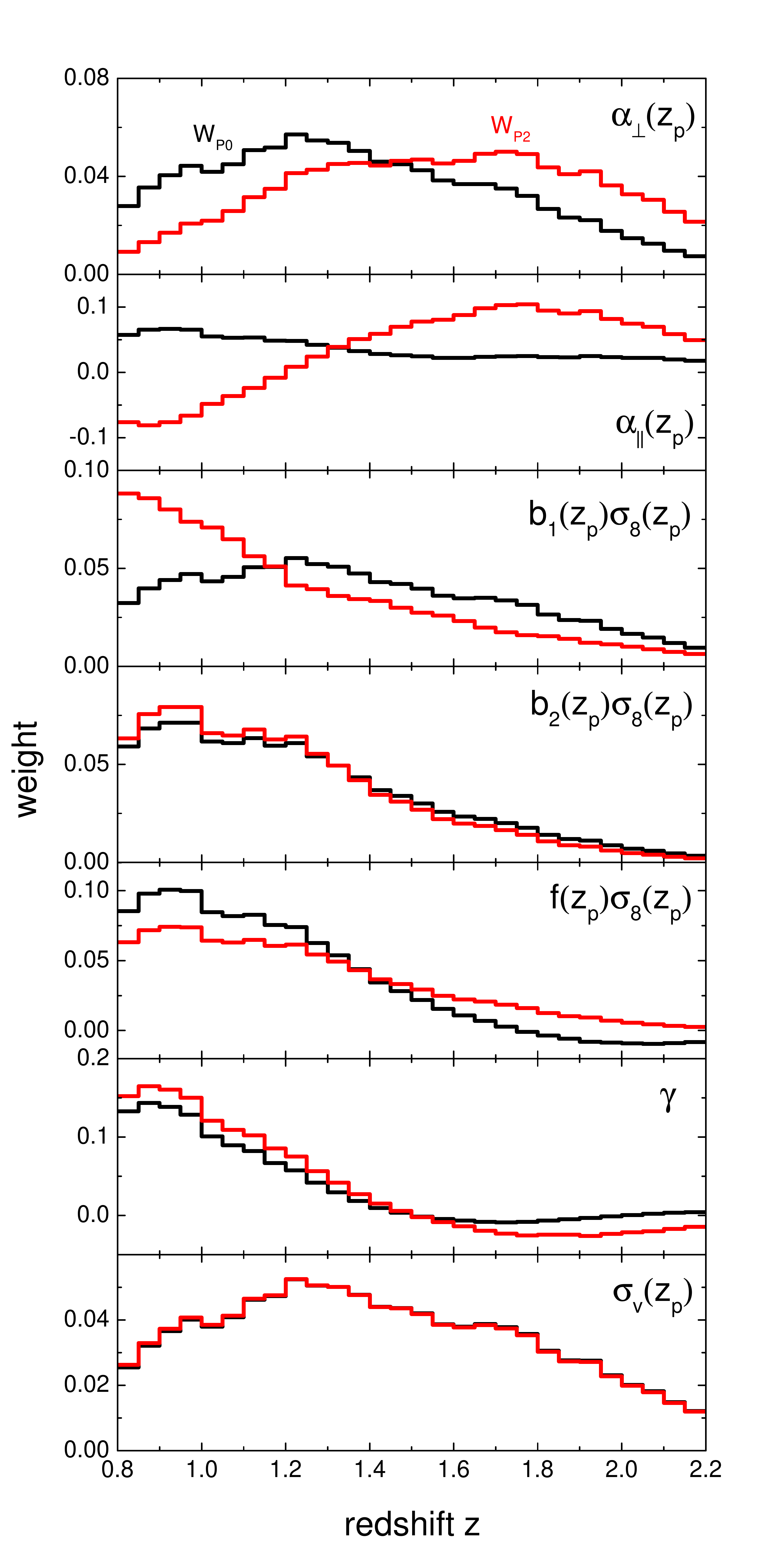}}
\caption{The optimal redshift weights for the monopole (black) and quadrupole (red) each free parameters as shown in the figure legend. For illustration, the weights are normalised so that the sum of each weight in the entire redshift range ($0.8<z<2.2$) is unity, although the normalisation can be arbitrary.}
\label{fig:weightp0p2}
\end{figure}

\begin{figure}
\centering
{\includegraphics[scale=0.32]{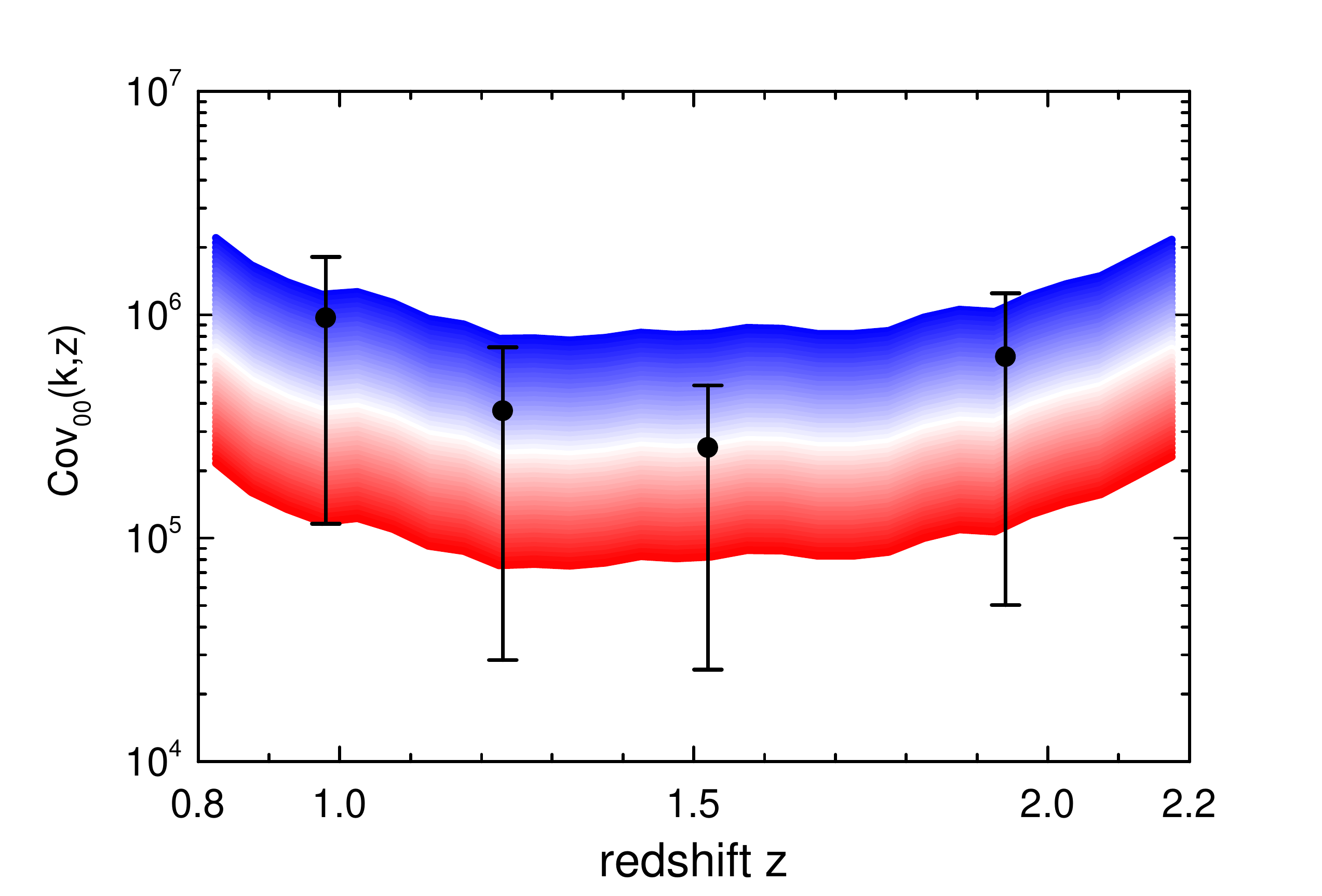}}
\caption{\tc A comparison of ${\rm Cov}_{00}(k,z)$, which are the diagonal elements of the covariance matrix (the monopole-monopole block), derived from the EZ mocks (data points with error bars), with that computed using the analytical formula shown in Eq (\ref{eq:C}) (the filled band). Both the band and the error bars show ${\rm Cov}_{00}(k,z)$ for $k$-modes in the range of $k\in[0.05,0.15]\ \kunit$. For the analytic covariance shown in the band, $k$ is sampled logarithmically from $0.05$ to $0.15\ \kunit$ from top to bottom, and the white curve in the middle corresponds to $k=0.1 \ \kunit$. For the covariance derived from mocks at four effective redshifts, the $k$-bins are made uniform linearly in the same range, and the central value and the error bars denote values for $k=0.1 \ \kunit$, and the standard deviation, respectively.}
\label{fig:C00}
\end{figure}

\begin{figure*}
\centering
{\includegraphics[scale=0.46]{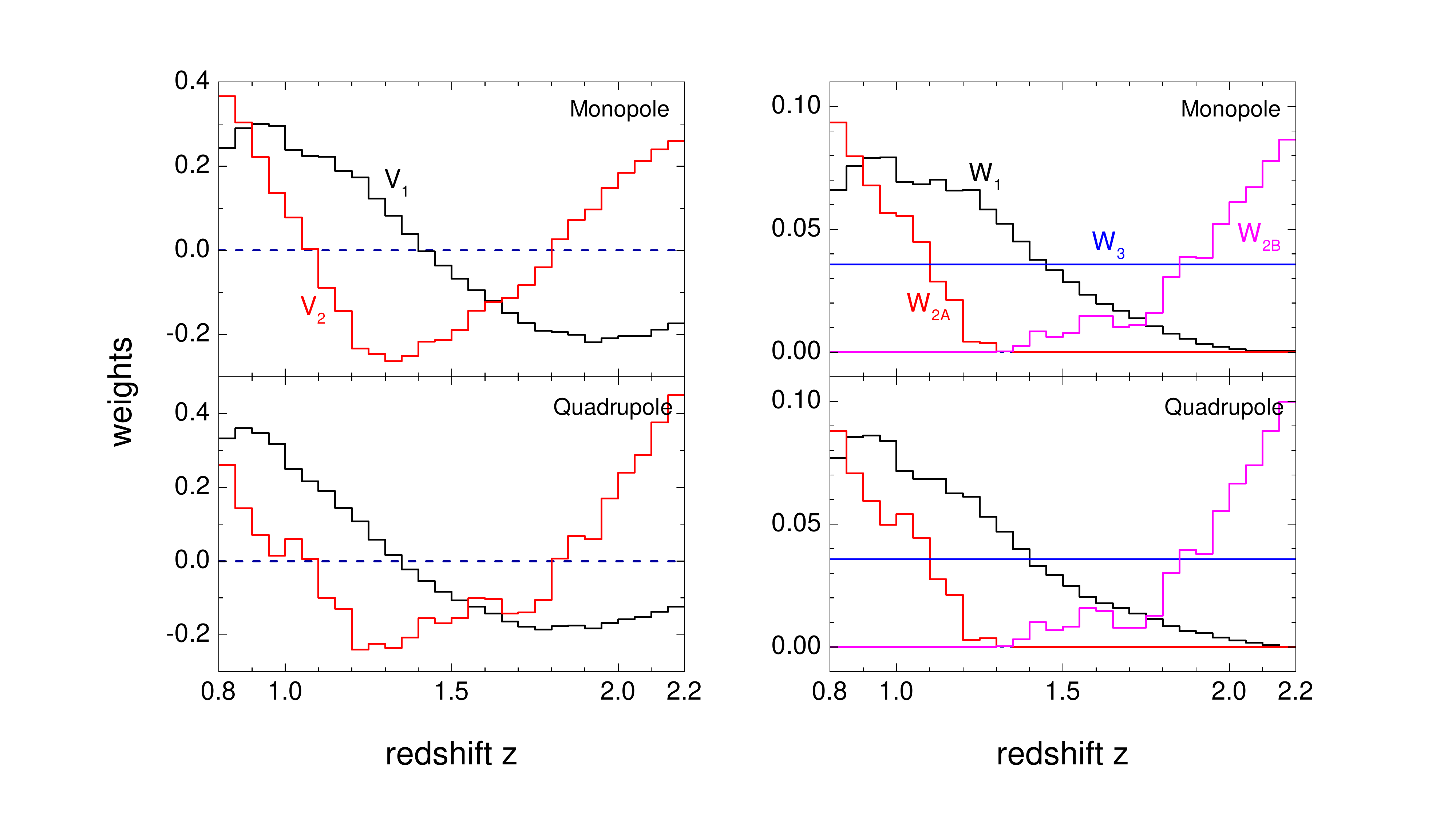}}
\caption{Left: The first two principal redshift weights, denoted as $V_1$ and $V_2$, for the monopole (upper panel) and quadrupole (lower) respectively derived from a SVD analysis. Right: The positive-definite redshift weights, $W_1, W_{\rm 2A(B)}, W_3$, derived from linear combinations of $V_1, V_2$ and a constant.
The $W$ weights are normalised in the same way as in Fig. \ref{fig:weightp0p2}.}
\label{fig:weightsvd}
\end{figure*}

\begin{figure*}
\centering
{\includegraphics[scale=0.6]{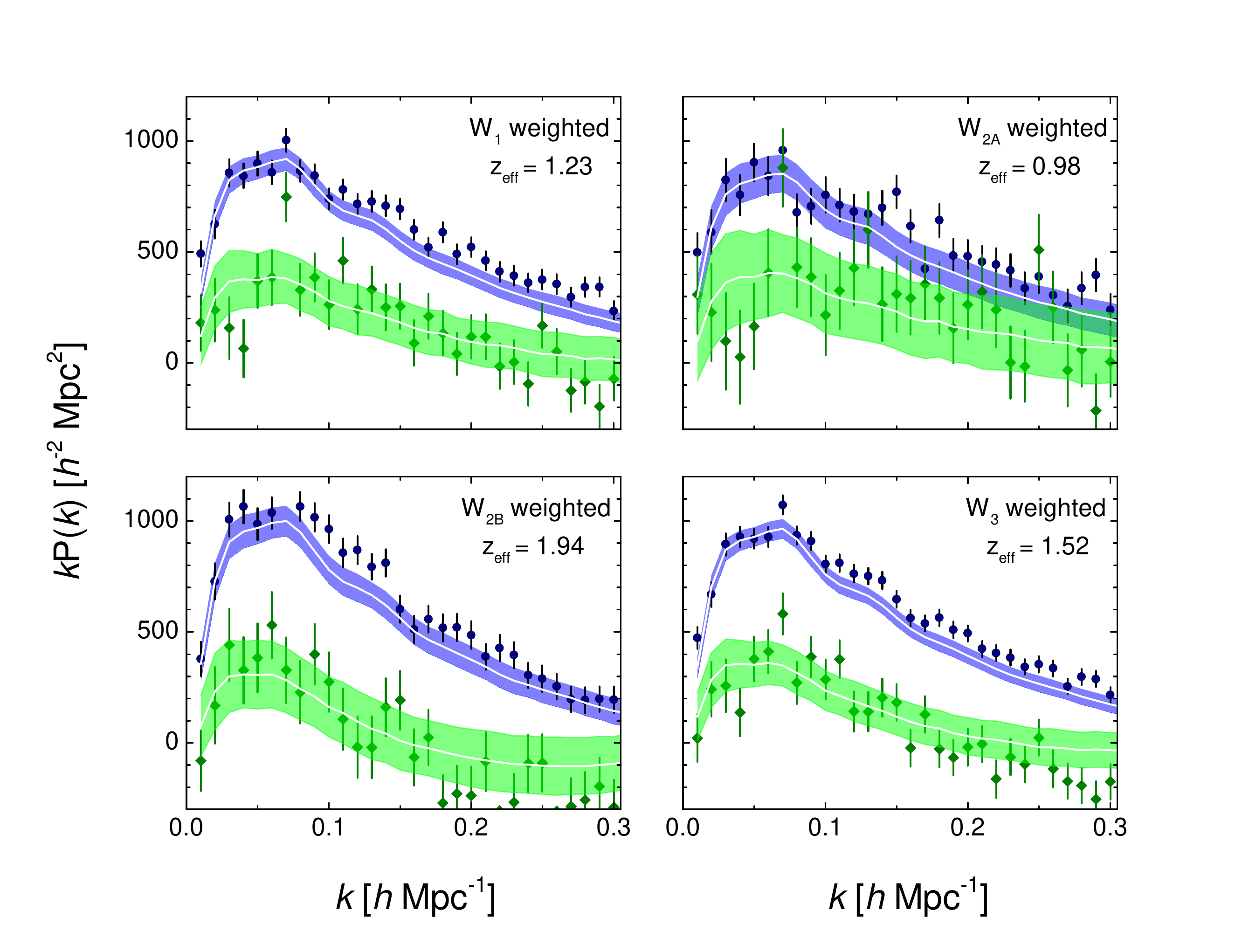}}
\caption{The measured galaxy power spectra monopole (upper band or data points) and quadrupole (lower) from the EZ mocks (filled bands) and DR14Q catalogues (data points with error bars) with redshift weights by $W_1, W_{\rm 2A}, W_{\rm 2B}$ and $W3$ respectively. All spectra are multiplied by the wavenumber $k$ for illustration.}
\label{fig:pkobs}
\end{figure*}

\begin{figure*}
\centering
{\includegraphics[scale=0.3]{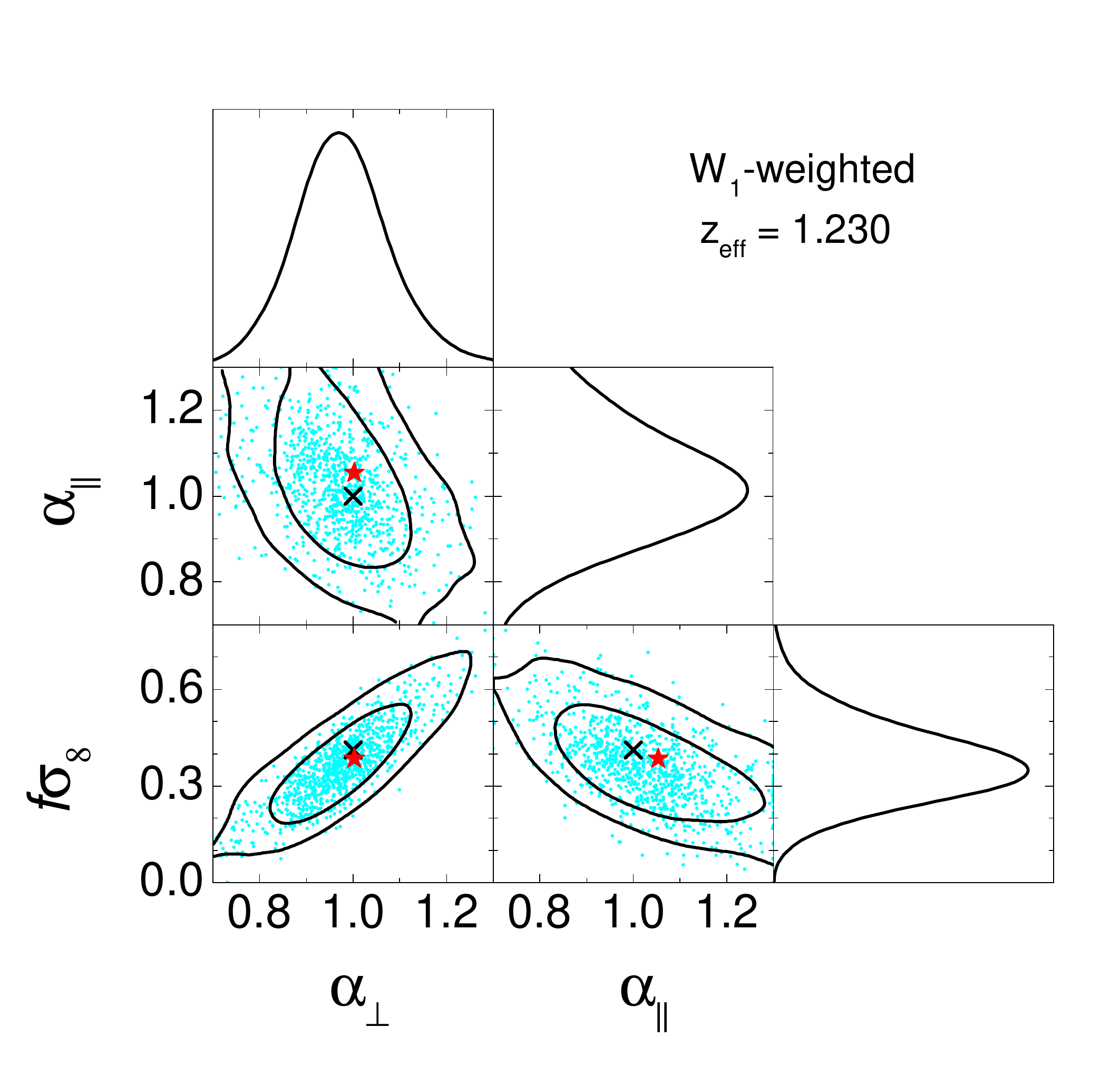}}
{\includegraphics[scale=0.3]{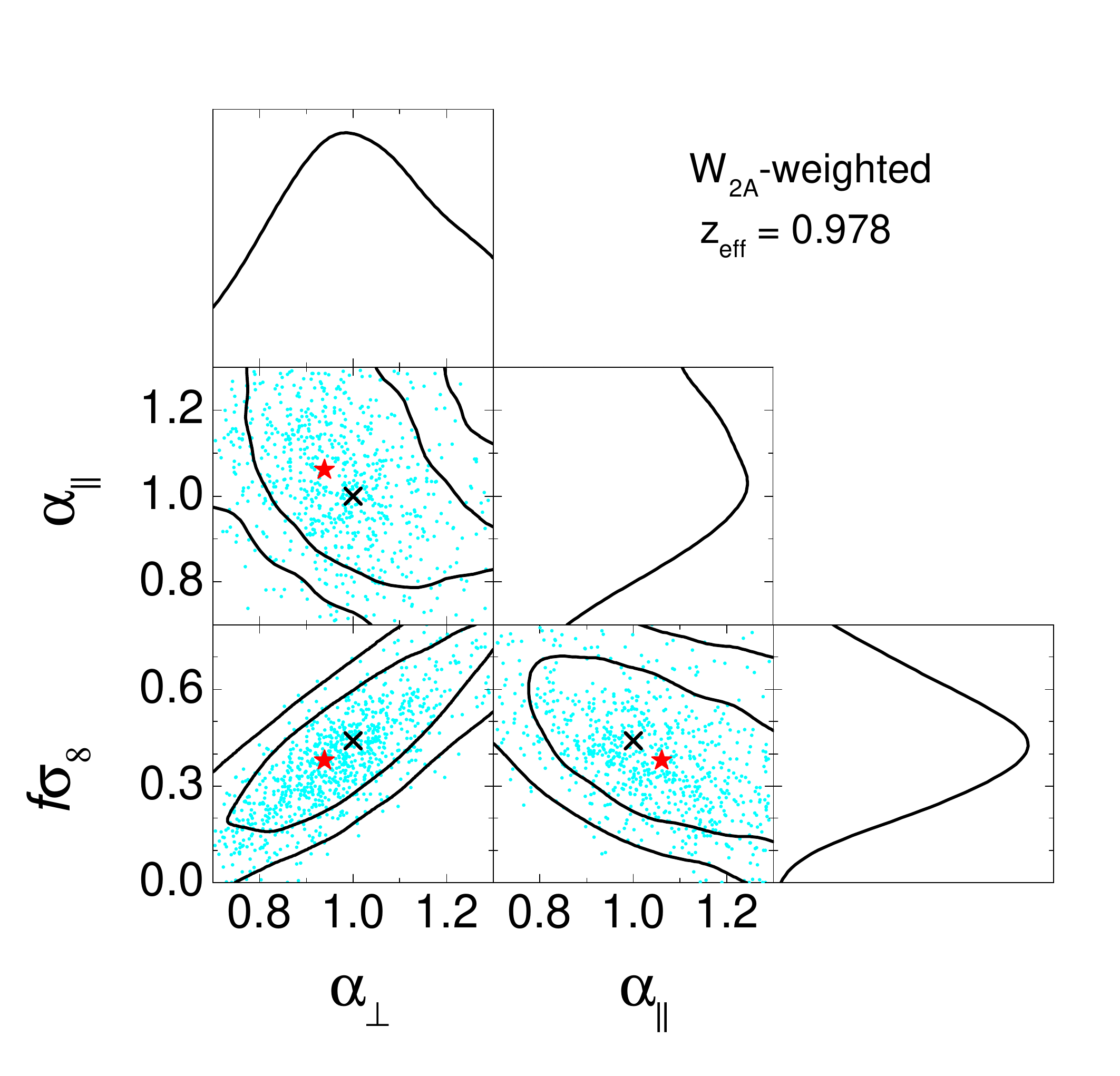}}
{\includegraphics[scale=0.3]{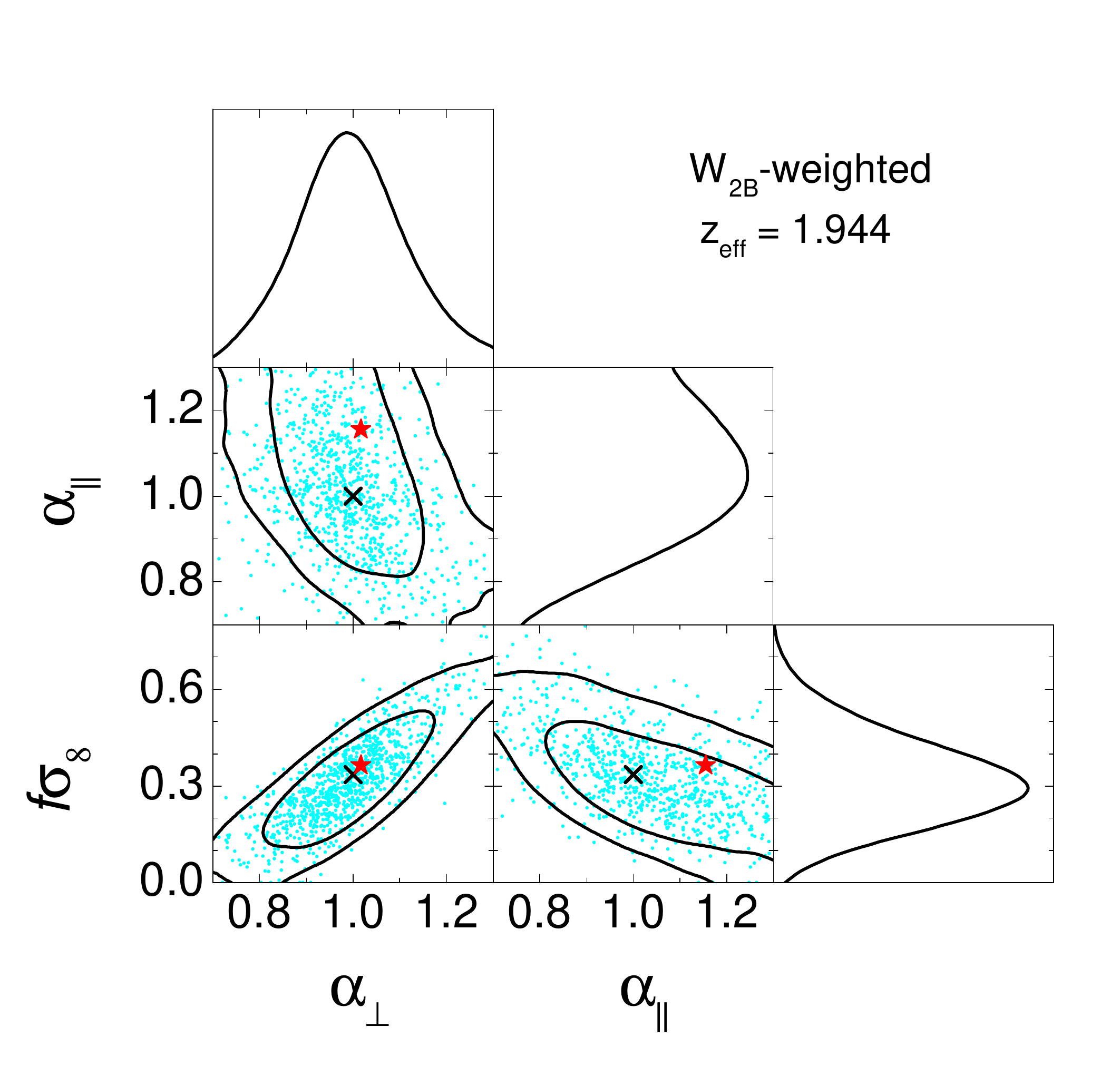}}
{\includegraphics[scale=0.3]{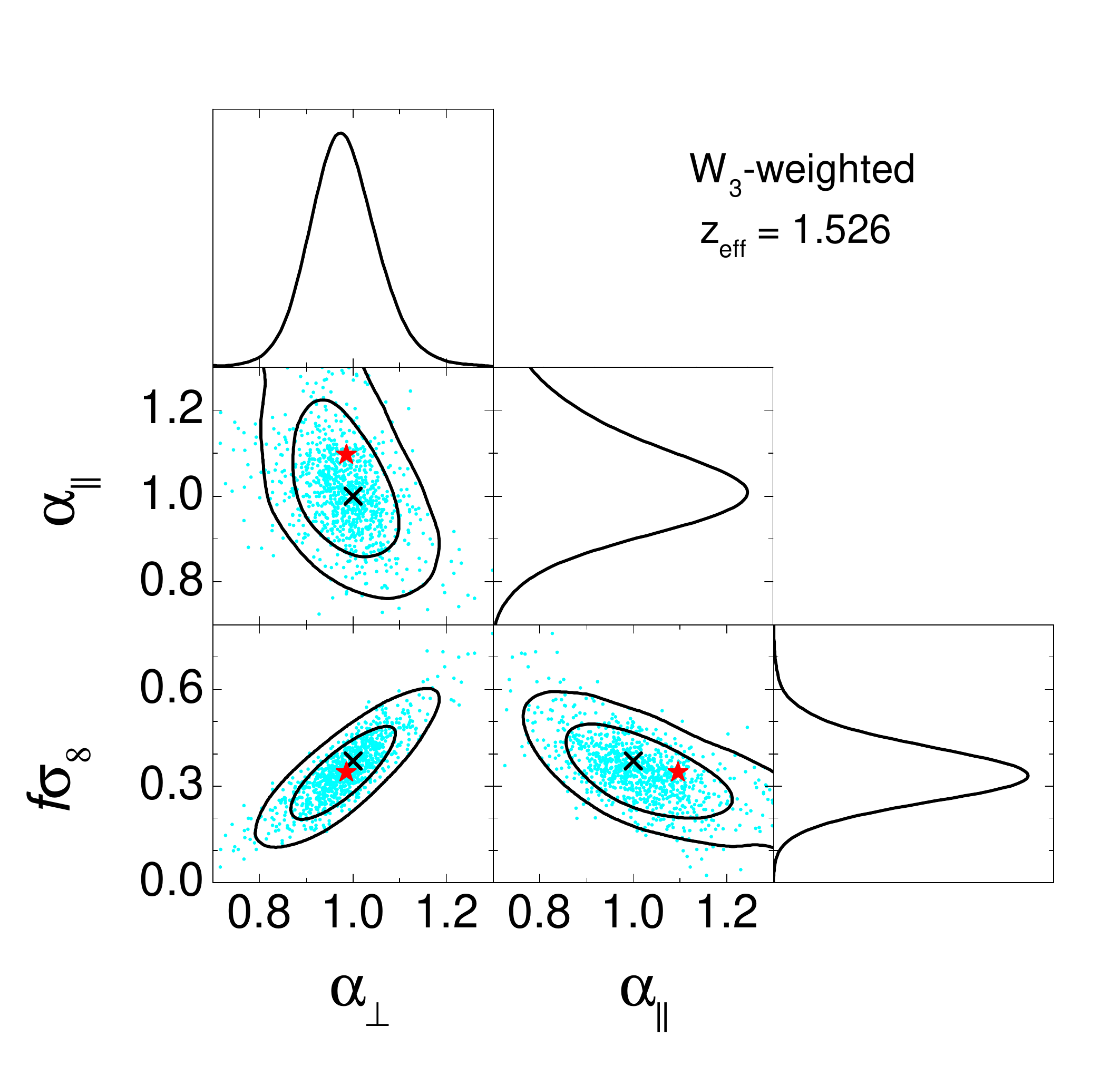}}
\caption{The result of mock tests and actual measurement of BAO and RSD parameters from the DR14Q sample. Results for four redshift-weighted samples are shown in four panels, as illustrated in the legend. In each panel, the best-fit values of parameters for each of the one thousand EZmocks are shown in cyan dots, and the black crosses and the red stars mark the expected values of EZmocks, and the actual measurement from the DR14Q sample respectively. The 68 and 95\% CL contours and one-dimensional posterior distribution of parameters are shown in black curves.}
\label{fig:mocks}
\end{figure*}

\begin{figure*}
\centering
{\includegraphics[scale=0.32]{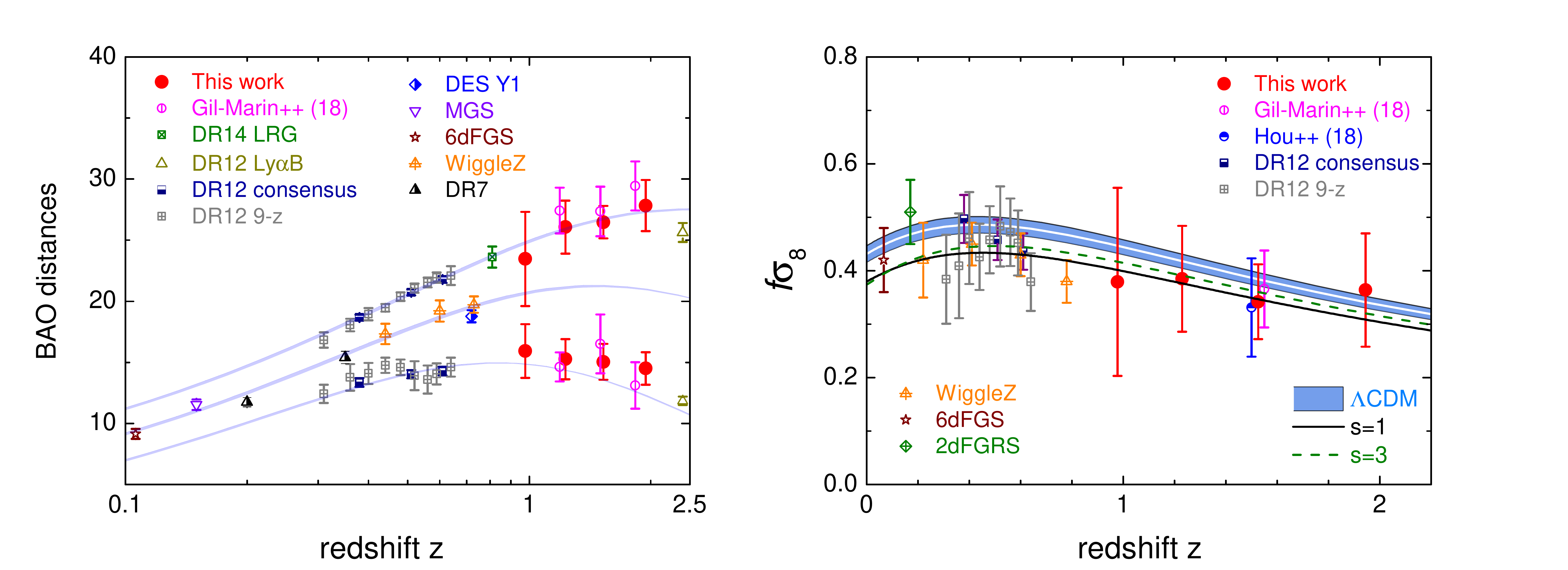}}
\caption{Left: The BAO distance measurements derived from this work presented in Tables \ref{tab:4z} and \ref{tab:corr} (large filled circles) in comparison with other recent BAO measurements (shown in the legend), including \citet{HGM18}, eBOSS DR14 LRG \citep{dr14lrg}, BOSS DR12 Ly$\alpha$F BAO \citep{dr12lyaf}, BOSS DR12 consensus \citep{alam}, BOSS DR12 tomographic BAO measurements at nine effective redshifts (DR12 9-$z$; \citealt{Zhaotomo16}), DES year 1 BAO \citep{desy1}, MGS \citep{mgs}, 6dFGS \citep{6dF}, WiggleZ BAO \citep{wigglezbao} and BOSS DR7 \citep{sdss2bao}. The three filled bands from top to bottom show the 95\% CL constraints of $D_{\rm M}(z)/\left(\rd\sqrt{z}\right)$, $D_{\rm V}(z)/\left(\rd\sqrt{z}\right)$ and $zD_{\rm H}(z)/\left(\rd\sqrt{z}\right)$ respectively. The band are derived from Planck 2015 observations, combined with external datasets including supernovae, galaxy clustering and weak gravitational lensing in a $\Lambda$CDM Universe \citep{ZhaoDE17}. The top and bottom bands and data points are vertically displaced by $2$ for illustration. Right: The RSD measurement parametrised by $f\sigma_8$ derived from this work (large filled circles) in comparison with other recent RSD measurements (shown in the legend), including \citet{HGM18}, \citet{Hou18}, BOSS DR12 consensus \citep{alam}, BOSS DR12 tomographic RSD measurements at nine effective redshifts (DR12 9-$z$; \citealt{Wangtomo17}), WiggleZ \citep{wigglezrsd}, 6dFGS \citep{6dfrsd} and 2dFGRS \citep{2dfrsd}. The filled band shows the mean, and 68\% CL constraint on $f\sigma_8$, derived from Planck 2015 observations, combined with external datasets in a $\Lambda$CDM Universe \citep{ZhaoDE17} (as in the left panel), {\tc and the black solid and green dashed curves show the best-fit modified gravity models denoted by two different values of $s$, which was derived in \citet{LZ18}. See texts for more details.}}
\label{fig:BAORSD}
\end{figure*}

\begin{figure*}
\centering
{\includegraphics[scale=0.35]{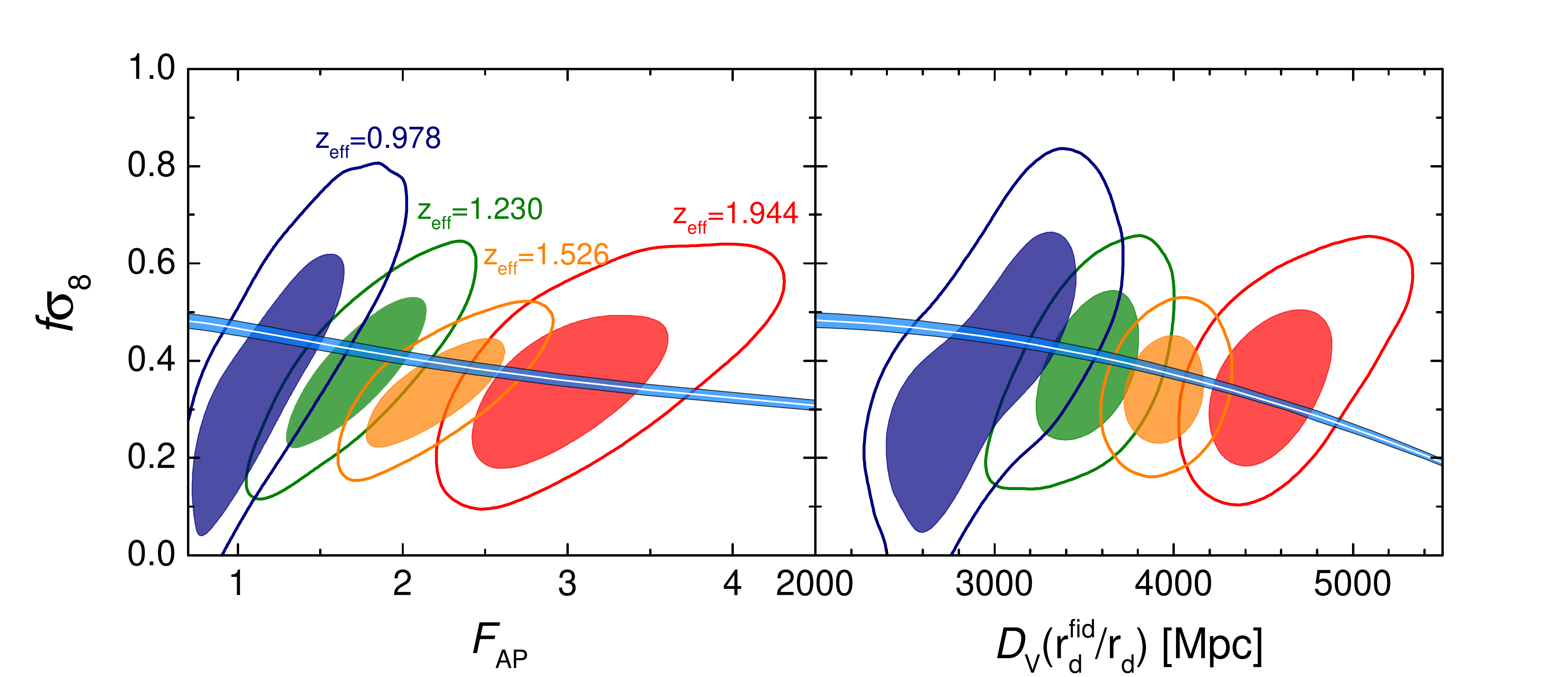}}
\caption{The 68 and 95\% CL contour plots between $f\sigma_8$ and $F_{\rm AP}$ (left panel), and between $f\sigma_8$ and $D_{\rm V}$ (right). In each panel, the contours from left to right are for measurements at four effective redshifts, as illustrated in the legend. As in the right panel of Fig. \ref{fig:BAORSD}, the filled bands show the mean, and 68\% CL constraint on $f\sigma_8$, derived from Planck 2015 observations, combined with external datasets in a $\Lambda$CDM Universe \citep{ZhaoDE17}.}
\label{fig:FAP_DV_fs8}
\end{figure*}

\begin{figure}
\centering
{\includegraphics[scale=0.36]{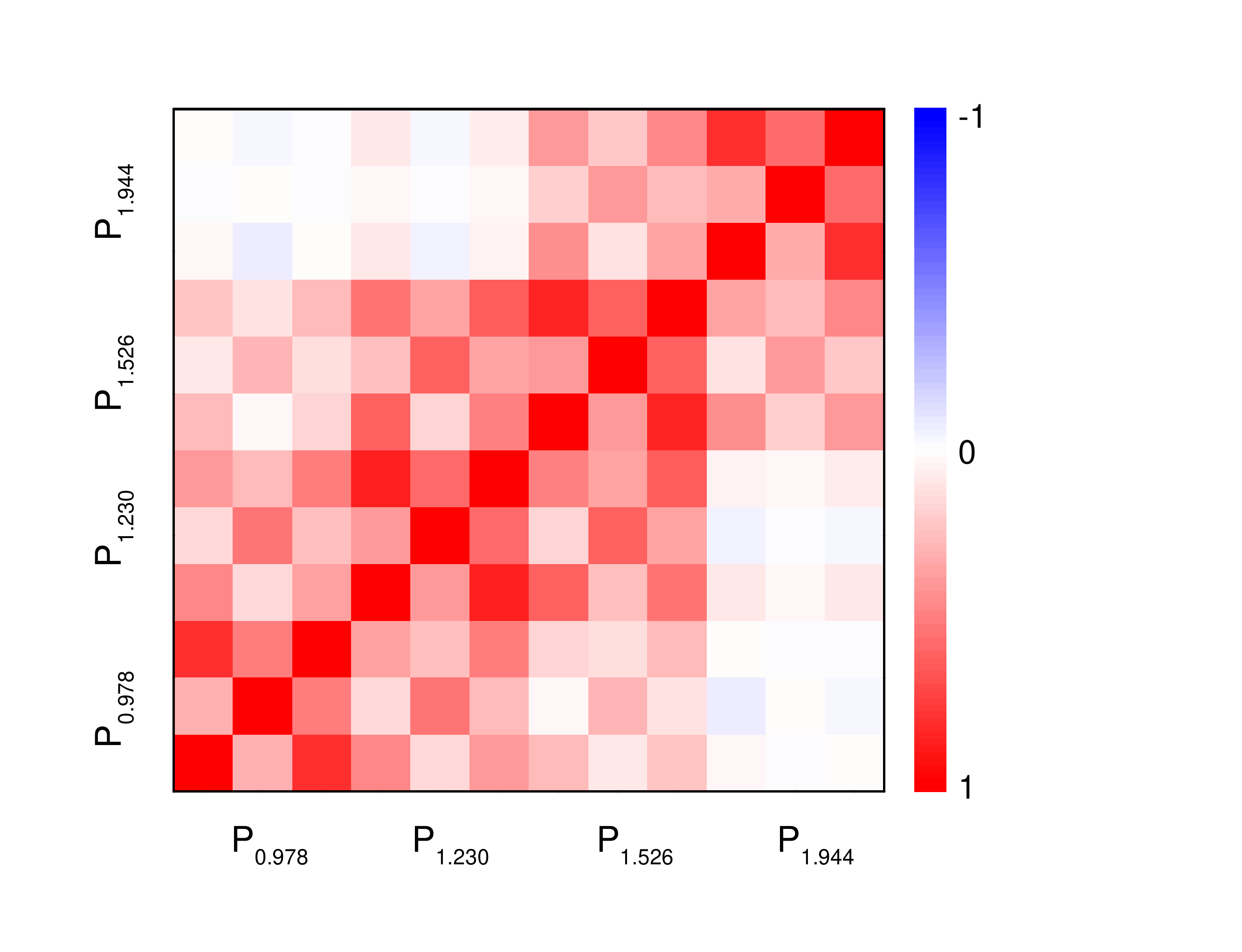}}
\caption{The correlation matrix for parameters measured from observables weighted by redshift weights shown in the legend. For each $3\times3$ parameter block, the order of parameters is $D_{\rm A}, H$ and $f\sigma_8$.}
\label{fig:corr}
\end{figure}

\begin{figure}
\centering
{\includegraphics[scale=0.35]{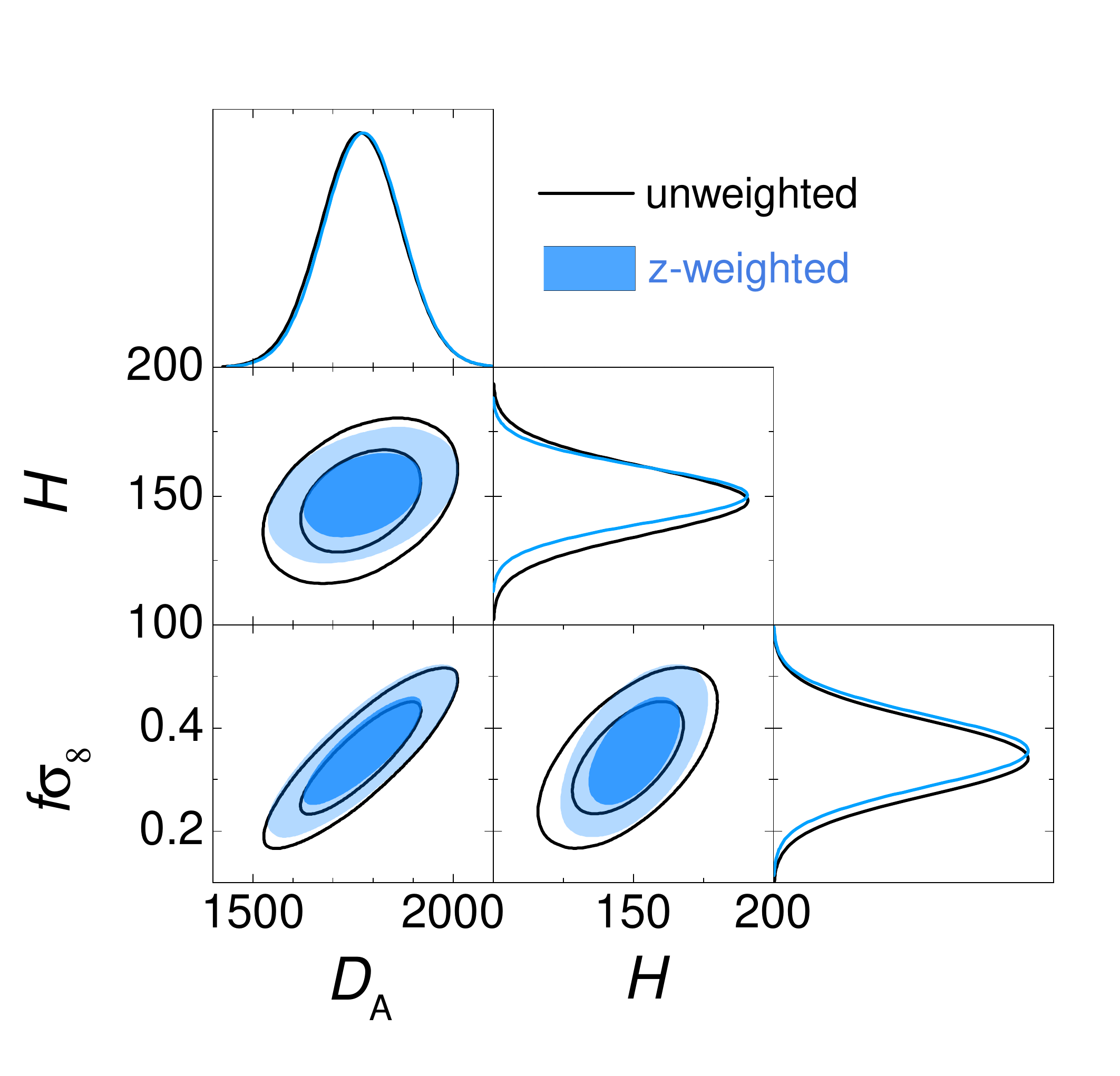}}
\caption{A comparison of the constraint on $D_{\rm A}$, $H$ and $f\sigma_8$ with (blue, filled) and without (black, unfilled) the redshift weights. The inner and outer contours show the 68 and 95\% CL two-dimensional marginalised constraints, and the one-dimensional curves show the posterior likelihood distribution for the corresponding parameters.}
\label{fig:contour}
\end{figure}

\begin{figure}
\centering
{\includegraphics[scale=0.36]{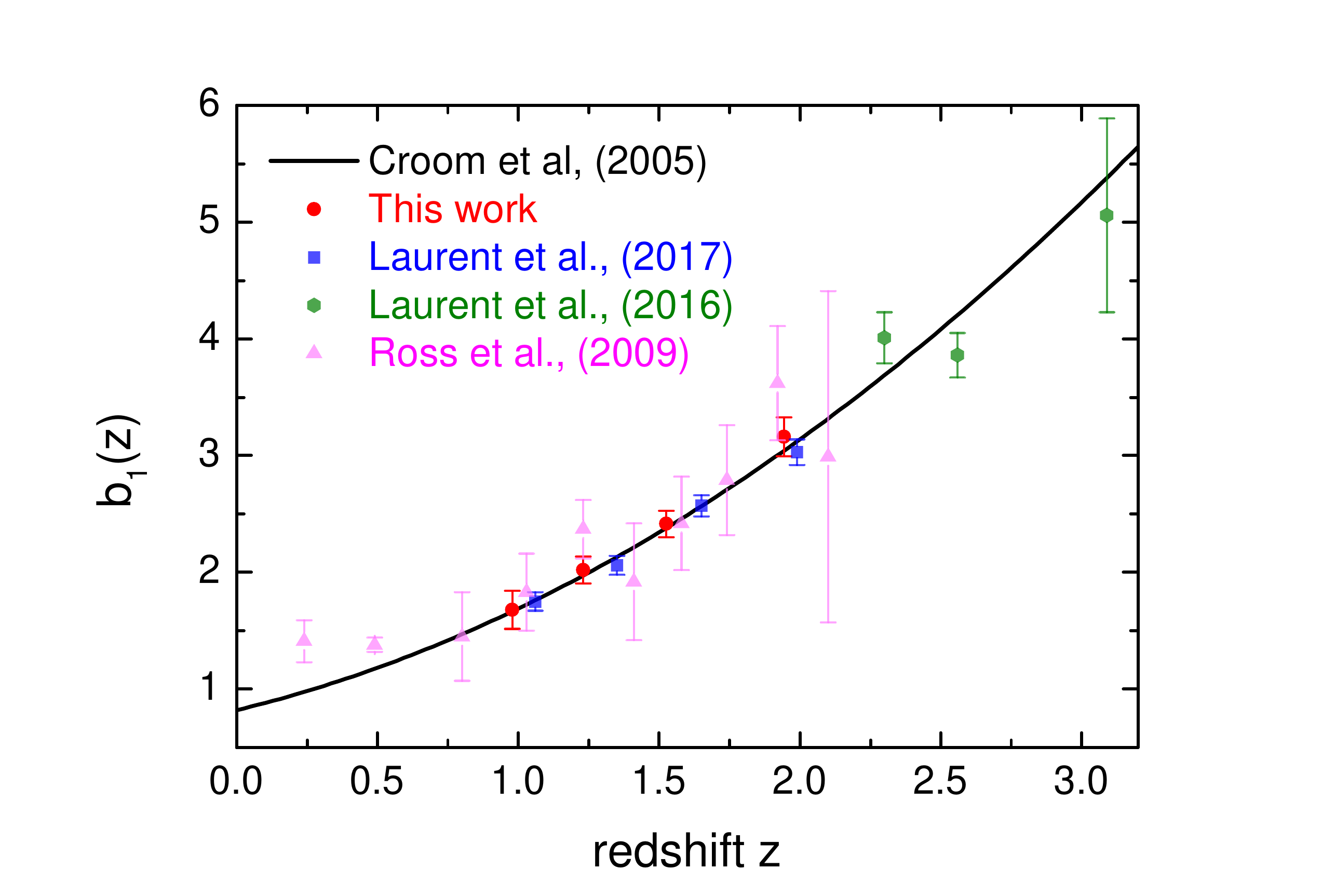}}
\caption{The measurement of the linear bias $b_1$ from this work (red circles), in comparison with other recent measurements denoted in the legend, including \citet{QSObias,Laurent16} and \citet{Ross09}. The black solid curve shows the model prediction of \citet{Croom}.}
\label{fig:bias}
\end{figure}

\subsection{The Alcock-Paczynski effect}
\label{sec:AP}

The Alcock-Paczynski (AP) effect quantifies the difference in the dilation of scales along and cross the line of sight due to the wrong cosmology used to convert redshifts to distances \citep{AP}, therefore this effect can be used to infer the true cosmology by contrasting the clustering along different lines of sight. Mathematically, the AP effect can be formulated as follows,

\ba\label{eq:Pell} P_{\ell}(k,z) = \frac{2\ell+1}{2\alpha_{\bot}^2\alpha_{||}}\int_{-1}^{+1}{\rm d}\mu \ P_g(k', \mu',z)\mathcal{L}_{\ell}(\mu) \ea where $P_g(k', \mu',z)$ is given by Eq (\ref{eq:Pgkmu}), $\mathcal{L}_{\ell}$ is {\tc the Legendre polynomial of order $\ell$}, and, 

\ba && k'  =  \frac{k(1+\epsilon)}{\alpha} \left\{1+\mu^2\left[\left(1+\epsilon\right)^{-6}-1\right]\right\}^{1/2}  \nonumber \\
   &&\mu' =  \frac{\mu}{(1+\epsilon)^3} \left\{1+\mu^2\left[\left(1+\epsilon\right)^{-6}-1\right]\right\}^{-1/2}    \nonumber \\
   && \alpha = \alpha_{\perp}^{2/3}\alpha_{\parallel}^{1/3}; ~~ 1+\epsilon = F_{\rm AP}^{1/3}; ~~F_{\rm AP}=\frac{\alpha_{\parallel}}{\alpha_{\perp}}
\ea The power spectrum monopole and quadrupole of the fiducial model at various redshifts from $0.8$ to $2.2$ are shown in Fig. \ref{fig:pkz}.

\subsection{The data covariance C}

We model the time evolution of the data covariance matrix {\bf C} using an analytic method \citep{TNS},

\ba \label{eq:C} && {\rm Cov_{\ell\ell'}}(k,z)=\frac{4\pi^2}{k^2\Delta k\Delta V(z)}\frac{(2\ell+1)(2\ell'+1)}{2}  \nonumber\\
 &&\times \int_{-1}^{+1}\d\mu\mathcal{L}_{\ell}(\mu) \mathcal{L}_{\ell'}(\mu)\left[P_g(k,\mu,z)+\frac{1}{\bar{n}_g(z)}\right]^2\ea 

{\tc In Figure \ref{fig:C00}, we show a comparison of ${\rm Cov}_{00}(k,z)$, the diagonal elements of the data covariance matrix (the monopole-monopole block), derived from the EZ mocks (the four data points with error bars), with that computed using the analytical formula shown in Eq (\ref{eq:C}) (the filled band). This shows that our analytic formula can well capture the redshift evolution of ${\rm Cov}_{00}$, especially at $k\simeq 0.1 \ \kunit$. We have numerically confirmed that this also holds for ${\rm Cov}_{22}$ and ${\rm Cov}_{02}$\footnote{We have checked and confirmed that the redshift evolution of ${\rm Cov}_{\ell\ell'}$ where $\ell, \ell'=0, 2$ shows a similarity to a large extent for different $k$-modes for $k\in[0.05,0.15]\ \kunit$.}. 

We notice that Eq (\ref{eq:C}) assumes the diagonality of the covariance among different $k$-modes, which well approximates the covariance matrices derived from the mocks, although it can be further improved using more sophisticated methods to include the non-Gaussian contribution (see \citealt{HP17,OE18} for recent developments and references therein).

Note that the amplitude of the estimated ${\rm Cov}$ is irrelevant, as long as the normalisation is kept the same for all redshifts. This is because the normalisation of the weights to be derived from ${\rm Cov}$ can be arbitrary.
}

\subsection{The optimal redshift weights}
\label{sec:zweight}

Now we attempt to derive the optimal redshift weights for each parameters {\tc shown in Table \ref{tab:param}.} Specifically, we evaluate the derivative matrix ${\bf D}$ shown in Eq (\ref{eq:D}) numerically, and compute the data covariance matrix ${\bf C}$ using Eq (\ref{eq:C}). We have numerically verified that the $k$-dependence of {\tc all} the weights is very weak in the $k$ range of $0.05\lesssim k \lesssim0.25\kunit$  where data are most informative, thus we compute the redshift weights at $k=0.1\kunit$ without loss of generality.

{\tc The optimal redshift weights for the relevant parameters using the monopole and quadrupole of the galaxy power spectrum are calculated using Eq (\ref{eq:opW}), and are shown in Fig. \ref{fig:weightp0p2}.} As illustrated, the shapes of these weights show a high level of similarity, which means that there would be significant redundancy in weighted power spectra, if these weights were applied. This will not only cause 
unnecessary computations, but also yield a largely singular data covariance matrix, which is difficult to invert accurately for likelihood analysis.   

{\tc To remove the redundancy in the redshift weights, we perform a singular-value decomposition (SVD) of the original redshift weights for all the parameters shown in Figure \ref{fig:weightp0p2}}, \ie,
\ba {\bf X} = {\bf U \Lambda V}^T \ea where ${\bf X}$ is the data matrix of the weights, and ${\bf \Lambda}$ is a diagonal matrix storing the variances. The new orthogonal weights can be found by projecting the original ones onto the new basis ${\bf V}$, whose variances are ordered in ${\bf \Lambda}$. Keeping a first few principal components can largely reduce the redundancy with negligible information loss \footnote{We provide a Matlab code in the Appendix for the SVD analysis.}.  

This procedure yields two orthogonal weights for monopole and quadrupole each, which represents over $90\%$ of the variance in the data, as shown in the left panels of Fig. \ref{fig:weightsvd}. Note that these new redshift weights are not generically positive definite, making it difficult to apply to the galaxy catalogues \footnote{This is because these weights are supposed to be applied to power spectra, not to individual galaxies. The weights for galaxies are square root of the redshift weights, thus they must be positive definite.}. In some occasions, all the weights can be made positive by a linear transformation without loss of information. However, as this is not always feasible, we add a third weight, which is a constant in $z$, to guarantee that the weights can be always turned positive by a linear transformation. As the added constant weight spoils the orthogonality of the weights, we tune the constant to minimise the correlation between the weights, which removes the redundancy as much as possible. {\tc We include a detailed procedure of obtaining the weights in Appendix \ref{sec:B}.} The resultant weights are shown in the right panel of Fig. \ref{fig:weightsvd}. 

Care must be taken when analysing these redshift-weighted samples using a template at a single effective redshift, as in the traditional method. As the redshift weights can be generally arbitrary in shape, it can make the redshift distribution of the weighted sample multi-modal, making it inaccurate 
to be modelled using a template at a single effective redshift. To be explicit, we revisit the calculation of the effective redshift.
Observationally, the measured power spectra are actually a redshift-weighted average across the redshift range of the catalogue, \ie,  
\ba\label{eq:p} P=\frac{\sum P(z_i) w_i^2}{\sum w_i^2} \ea
Expanding the power spectra at an arbitrary redshift $z$ around an effective redshift $\zeff$ yields,
\ba\label{eq:pz} P(z)=P(\zeff)+P'(z-\zeff)+\frac{1}{2}P''(z-\zeff)^2+\mathcal{O}(P''') \ea
Combining Eqs (\ref{eq:p}) and (\ref{eq:pz}), we have,
\ba P&=&P(\zeff) + P'\Delta_1 + \frac{1}{2}P''\Delta_2 +\mathcal{O}(P''') \ea
where,
\ba \Delta_1&=&\left(\frac{\sum z_i w_i^2}{\sum w_i^2} -\zeff\right)  \nonumber \\
\Delta_2&=&\left(\frac{\sum z_i^2 w_i^2}{\sum w_i^2} -2\zeff \frac{\sum z_i w_i^2}{\sum w_i^2}+z^{2}_{\rm eff} \right) \nonumber \\ 
\ea
The first-order term $\Delta_1$ vanishes if $\zeff=\frac{\sum z_i w_i^2}{\sum w_i^2}$, but this does not necessarily  diminish $\Delta_2$ and higher order terms. Actually, when $\Delta_1=0$,

\ba\label{eq:d2} \Delta_2&=&\left[\frac{\sum z_i^2 w_i^2}{\sum w_i^2} -\left(\frac{\sum z_i w_i^2}{\sum w_i^2}\right)^2\right]  \ea
The catalogue can only be analysed using a template at the effective redshift if $\Delta_2\ll1$, which is not always the case generally. We have numerically checked that $\Delta_2(W_1)$ and $\Delta_2(W_3)$ are sufficiently small to be ignored. However, this term for $W_2$ ($W_{\rm 2A}+W_{\rm 2B}$ in the right panel of Fig. \ref{fig:weightsvd}) is non-negligible due to its double-peaked feature. Therefore we split this weight into two pieces $W_{\rm 2A}$ and $W_{\rm 2B}$ so that each one can be well modelled by its own effective redshift. The explicit values for $\Delta_2$ for these four weighted samples are listed in the bottom of Table \ref{tab:4z}.

\section{Results}
\label{sec:result}
In this section, we perform tests on the mocks before the joint measurement on BAO and RSD parameters using the eBOSS DR14Q sample at four effective redshifts. We also present a measurement of linear bias.

\subsection{Joint BAO and RSD measurements}
\label{sec:BAORSD}

We first apply the square root of redshift weights $W_1, W_{\rm 2A}, W_{\rm 2B}$ and $W_3$ shown in Fig. \ref{fig:weightsvd} to both the EZ mocks and the DR14Q sample, and measure the corresponding power spectra monopole and quadrupole using the FFT method presented in \cite{PkFFT}, as shown in Fig. \ref{fig:pkobs}. As the power spectra derived from the $z$-weighted samples are essentially linear combinations of power spectra at multiple redshifts, we compute the effective redshifts for each of the weighted sample using Eq (\ref{eq:zeff}), and find, \ba z_{\rm eff}(W_1) = 1.23, \nonumber \\
      z_{\rm eff}(W_{\rm 2A}) = 0.98, \nonumber \\
      z_{\rm eff}(W_{\rm 2B}) = 1.94, \nonumber \\      
      z_{\rm eff}(W_3) = 1.52.  \ea
      
Using a modified version of CosmoMC \citep{cosmomc}, we then fit for parameters shown in Sec. \ref{sec:param} at each effective redshift to the power spectra using the template detailed in Sec. \ref{sec:temp}. The theoretical power spectra multipoles are convolved with the survey window functions, which are shown in Sec. \ref{sec:window}, measured using the method developed in \cite{pkmask}. The joint measurement of $\abot,\apar$ and $\fs$ is shown in Fig. \ref{fig:mocks}. 

Each of the cyan dots represents the best-fit model derived from one specific EZ mock, and the black contours show the 68 and 95\% CL constraint using the mean of the 1000 EZ mocks. As shown, the measurement of BAO and RSD parameters (with other parameters marginalised over) at four effective redshifts are all largely consistent with the expected values denoted by black crosses with the maximal deviation less than $0.3\sigma$, which validates our pipeline.

We then apply our pipeline to the DR14Q catalogue, and show the measurement in Table \ref{tab:4z} and in Figs. \ref{fig:mocks}, \ref{fig:BAORSD} and \ref{fig:FAP_DV_fs8}.

In Fig. \ref{fig:mocks}, we see that the best-fit model to the DR14Q sample (red stars) is within the 68\% CL contours of the EZ mocks at all effective redshifts, which means that the fiducial cosmology used to produce the EZ mocks can reasonably approximate the true cosmology probed by the quasar sample within 68\% CL.

Fig. \ref{fig:BAORSD} shows our BAO and RSD measurement in comparison with other published ones from galaxy surveys, as well as with the Planck constraint in a $\Lambda$CDM model derived in \cite{ZhaoDE17}. Our measurements of $D_{\rm M}\equiv(1+z)D_{\rm A}$ and $\fs$ are in excellent agreement with the Planck constraint at all redshifts, but our measurement on $D_{\rm H}(z)\equiv c/H(z)$ shows a deviation at $z=1.526$ and $z=1.944$ at $\gtrsim 1 \sigma$ significance. Interestingly, \cite{HGM18} finds a similar deviation at $z=1.50$ using the same data sample. Moreover, the $D_{\rm H}$ measurement at $z=2.4$ using Lyman-$\alpha$ forest shows a deviation in the same direction. We will reinvestigate this issue when the eBOSS quasar survey is completed. 

As shown in the right panel of Fig. \ref{fig:BAORSD}, our $\fs$ measurement at $z=1.52$ is largely consistent with that presented in companion papers of \cite{HGM18} and \cite{Hou18}, which are studies on the same data sample using different methods. {\tc Interestingly, the compilation of $\fs$ measurements shown in the right panel of Figure \ref{fig:BAORSD} seems to favour lower values of $\fs$ than that in the $\Lambda$CDM model across a wide redshift range. \citet{LZ18} performed a constraint on modified gravity models using a combined observational data including the BAO and RSD measurement derived in this work, and it is found that a model in which the effective Newton's constant is parametrised as $G_{\rm eff} = 1+\mu_s a^s$ (where $\mu_s$ and $s$ are free parameters, and $\mu_s=0$ in $\Lambda$CDM) is able to fit the data better (see our overplot of their best-fit model predictions with data points in the right panel of Fig. \ref{fig:BAORSD}).}

We show contour plots between $\fs, D_{\rm V}$ and $F_{\rm AP}\equiv D_{\rm M}H/c$ in Fig. \ref{fig:FAP_DV_fs8}, and as shown, our measurements are consistent with the Planck observations.

As the weights are not orthogonal to each other, the measurements at four effective redshifts are generally correlated. We quantify the correlation by fitting to each of the 1000 EZ mocks, and compute the correlation matrix using the fitted parameters. The correlation matrix is shown in Fig. \ref{fig:corr}, with the numeric values of the correlation matrix and the precision matrix shown in Table \ref{tab:corr} \footnote{The correlation matrix and the precision matrix are the rescaled covariance matrix and inverse covariance matrix respectively, with all the diagonal elements being unity.}. As expected, the same parameters at different effective redshifts are positively correlated except for those at $z=0.978$ and at $z=1.944$, as the quasar distributions for these two weighted samples do not overlap.

Tables \ref{tab:4z} and \ref{tab:corr} present the main result of this work, which can be directly used to constrain cosmological models. To compare with measurements at $z_{\rm eff}=1.52$ presented in companion papers, we linearly combine our measurements at four redshifts. We adjust the coefficients for the combination so that the effective redshift calculated using Eq (\ref{eq:zeff}) is exactly $1.52$. Given that the set of coefficients to yield $\zeff=1.52$ is not unique, we choose a set of coefficients to maximise the FoM of $D_{\rm A}, H$ and $\fs$, with a constraint of $\Delta_2\ll1$ for the linearly combined sample. {\tc The procedure is explicitly shown in Appendix \ref{sec:A}.}

We have numerically checked that as long as $\Delta_2\ll1$ and the FoM saturates to its maximal value, different choices of the coefficients have negligible impact on the resultant parameter constraints. With these constraints, the coefficients are found to be $\{0.02, 0.17, 0.57, 0.24\}$ for the weighted samples with $z_{\rm eff}=0.978, 1.230, 1.526, 1.944$ respectively. {\tc Note that due to the correlation among the four catalogues, the trivial solution of $\{0, 0, 1, 0\}$ does not maximise the FoM (see Figure \ref{fig:contour_c} in the Appendix).}

The final measurement at $z_{\rm eff}=1.52$ is shown in Table \ref{tab:zweight} and in Fig. \ref{fig:contour}. To distinguish this measurement from the raw measurement at $z_{\rm eff}=1.52$, we denote this and the raw measurement as "$z$-weighted" and "unweighted" respectively. As shown, the "$z$-weighted" constraint is slightly tighter, namely, the FoM of $D_{\rm A}, H$ and $\fs$ is improved by 15\%. However, we strongly recommend users to use the tomographic measurement shown in Tables \ref{tab:4z} and \ref{tab:corr} for model constraints as those are more informative with lightcone information.

\input{tables/widetable1.tex}
\label{tab:4z}
\end{table*}%

\input{tables/widetable2.tex}
\label{tab:corr}
\end{table*}%

\input{tables/widetable3.tex}
\label{tab:zweight}
\end{table*}%

\subsection{A measurement of the linear bias}
\label{sec:bias}

As a by-product of our BAO and RSD measurements, we measure the linear bias $b_1$ at four effective redshifts, and present the result in Table \ref{tab:4z} and in Fig. \ref{fig:bias}. In Fig. \ref{fig:bias}, we overplot our measurement with published results using clustering quasars including \citet{QSObias,Laurent16,Ross09}, as well as with the fitting formula developed in \citet{Croom}. We find an excellent agreement between our measurement and the \citet{Croom} fitting formula \footnote{\tc We also measured the bias evolution from the EZ mocks, and find an excellent agreement with the \citet{Croom} fitting formula as well.}.

\section{The consensus result}
\label{sec:consensus}

The joint BAO and RSD analysis presented in this work is based on a power spectrum analysis using monopole and quadrupole (in the $k$-range of $0.02\leq k\,[h{\rm Mpc}^{-1}]\leq 0.30$) derived from the eBOSS DR14 quasar sample covering the redshift range of $0.8\leq z \leq 2.2$. The power spectrum template used in this work is primarily based on the regularised perturbation theory up to second order. With the optimal redshift weights, we constrain $D_{\rm A}, H$ and $f\sigma_8$ at four effective redshifts, namely, $z_{\rm eff}=0.978, 1.230, 1.562$ and $1.944$.

This work is released along with other complementary RSD analyses based on the exact same sample, including the same weighting schemes described in \cite{HGM18} (except for the redshift weights used in this work). The fiducial cosmology in which the sample has been analysed is also the same across papers. We briefly describe them below.

\begin{itemize}

\item The RSD analysis in \cite{HGM18} is based on the eBOSS DR14 quasar sample in the redshift range $0.8\leq z \leq 2.2$, using the power spectrum monopole, quadrupole and hexadecapole measurements on the $k$-range, $0.02\leq k\,[h{\rm Mpc}^{-1}]\leq 0.30$, shifting the centres of $k$-bins by fractions of $1/4$ of the bin size and averaging the four derived likelihoods. Applying the TNS model along with the 2-loop resumed perturbation theory, we are able to effectively constrain the cosmological parameters $f\sigma_8(z)$, $H(z)r_s(z_d)$ and $D_A(z)/r_s(z_d)$ at the effective redshift $z_{\rm eff}=1.52$, along with the remaining `nuisance' parameters, $b_1\sigma_8(z)$, $b_2\sigma_8(z)$, $A_{\rm noise}(z)$ and $\sigma_P(z)$, in all cases with wide flat priors. 

\item \cite{Hou18} analyses the eBOSS DR14 quasar sample in the redshift range $0.8\leq z \leq 2.2$ using Legendre polynomial with order $\ell=0,2,4$ and clustering wedges. They use "gRPT" to model the non-linear matter clustering. As for the RSD, they use a streaming model extended to one-loop contribution developed by \cite{SR04,TNS} and a nonlinear corrected FoG term. They adopt the bias modelling as described in \cite{Chan12}, which includes both local and nonlocal contribution.  Additionally they also include the modelling for spectroscopic redshift error. Finally they arrive at constraints on $f\sigma_8(z_{\rm eff})$ $D_{\rm V}(z)/r_{\rm d}$, $F_{\rm AP}(z)$ at the effective redshift $z_{\rm eff}=1.52$.

\item The clustering analysis presented in \cite{PZ18} is based on the eBOSS DR14 quasar sample in the redshift range $0.8\leq z \leq 2.2$, using Legendre multipoles with $\ell=0,2,4$ and three wedges of the correlation function on the $s$-range from 16~$h^{-1}{\rm Mpc}$ to 138~$h^{-1}{\rm Mpc}$. They use the Convolution Lagrangian Perturbation Theory (CLPT) with a Gaussian Streaming (GS) model and they demonstrate its applicability for dark matter halos of masses of the order of $10^{12.5}{\rm M}_\odot$ hosting eBOSS quasar tracers at mean redshift $z\simeq1.5$ using the OuterRim simulation. They find consistent results between the two methods and it yields to constraints on the cosmological parameters $f\sigma_8(z_{\rm eff})$, $H(z_{\rm eff})$ and $D_A(z_{\rm eff})$ at the effective redshift $z_{\rm eff}=1.52$.

\item  \cite{RR18} measures the growth rate and its evolution using the anisotropic clustering of the extended Baryon Oscillation Spectroscopic Survey  (eBOSS) Data Release 14 (DR14) quasar sample. To optimise the measurements we deploy a redshift-dependent weighting scheme, which avoids binning, and perform the data analysis consistently including the redshift evolution across the sample. They perform the analysis in Fourier space, and use the redshift evolving power spectrum multipoles to measure the redshift space distortions parameter $f\sigma_8$ alongside nuisance parameters, and parameters controlling the anisotropic projection of  the cosmological perturbations. They make use of two different sets of weights, described in \cite{zwRSD}. This model ties together growth and geometry, but can also be used after fixing the expansion rate to match the prediction of the $\Lambda$CDM model. The second parametrizes the $f\sigma_8$ parameter combination measured by RSD, allowing for a more standard test of deviations from $\Lambda$CDM. They compare all results with the \textit{standard} analysis performed at one single redshift of $z = 1.52$. 

\end{itemize}

\begin{figure}
\centering
\includegraphics[scale=0.55]{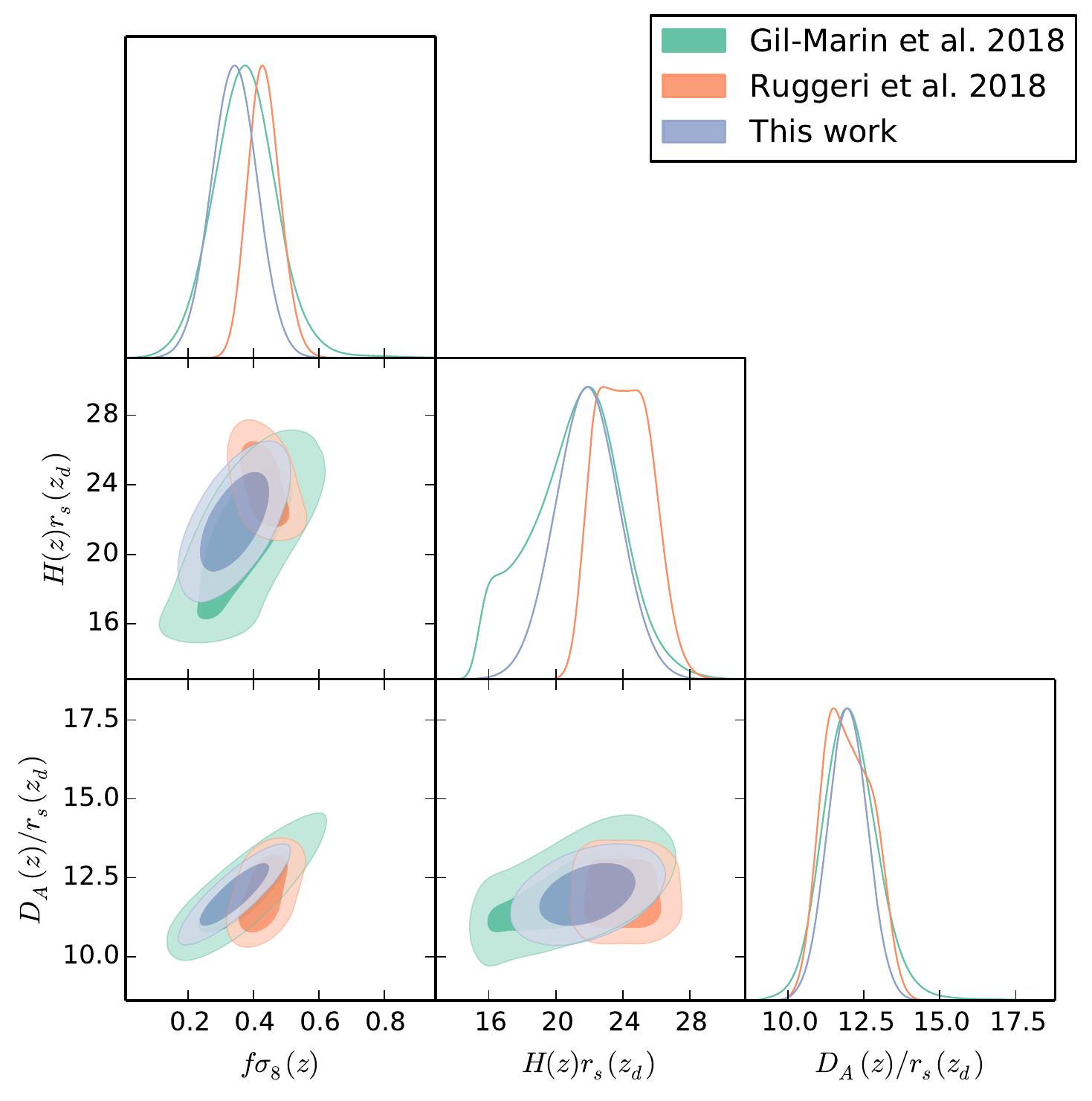}
\caption{Constraints on $f\sigma_8$, $D_A$ and $H$, in comparison with another two analysis in Fourier space, \citealt{HGM18} (green) and \citealt{RR18} (orange). No redshift weights are applied in all the analysis shown in this plot.}
\label{fig:consensusnoz}
\end{figure}

\begin{figure}
\centering
\includegraphics[scale=0.55]{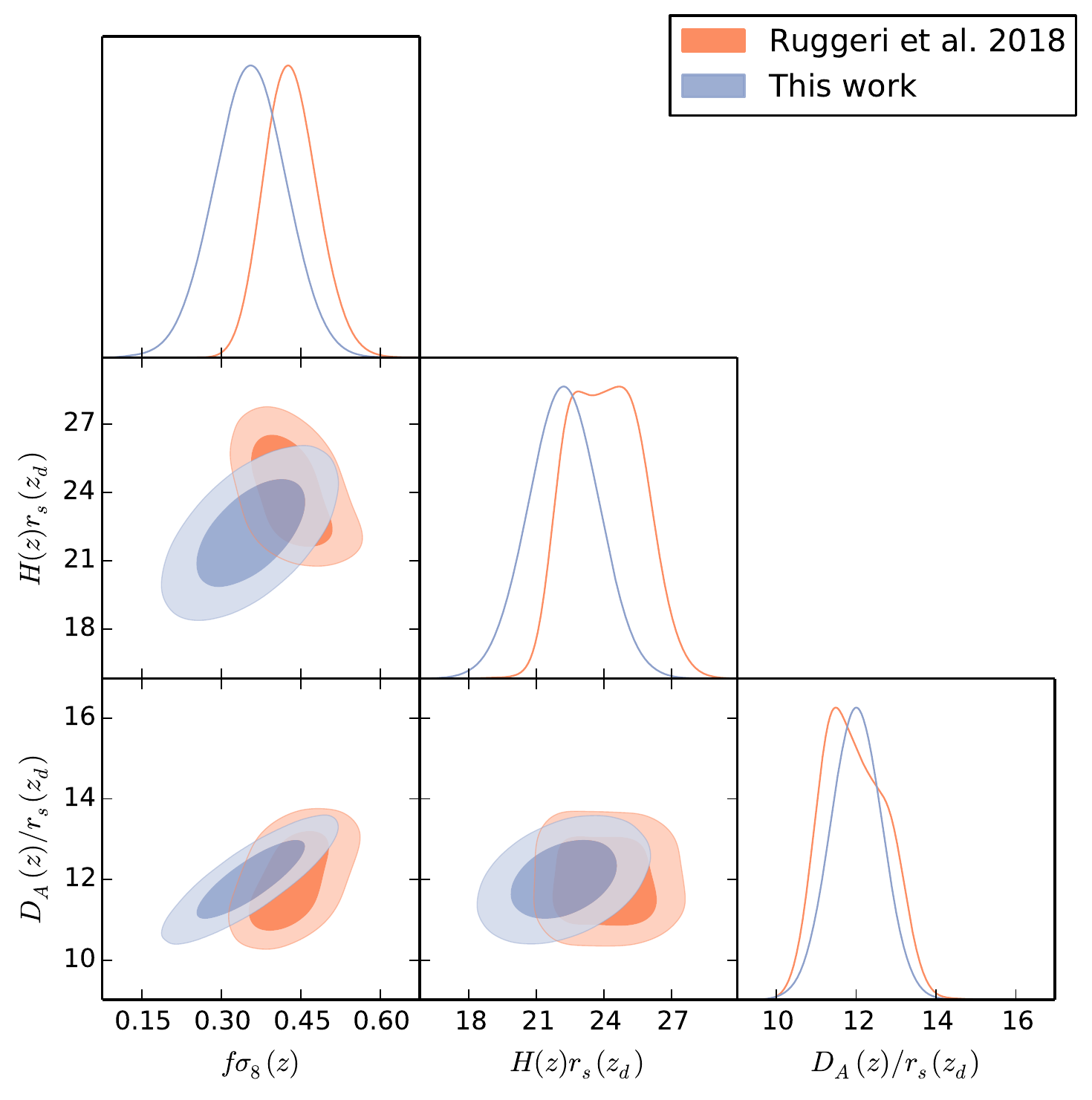}
\caption{Constraints on $f\sigma_8$, $D_A$ and $H$, in comparison with another analysis in Fourier space, \citealt{RR18} (orange). Redshift weights are applied in the analysis shown in this plot.}
\label{fig:consensusz}
\end{figure}

In Figs. \ref{fig:consensusnoz} and \ref{fig:consensusz}, we make a direct comparison to two of the companion works, which are RSD analysis in Fourier space. Results shown in Fig. \ref{fig:consensusnoz} are without redshift weights, while results in \ref{fig:consensusz} are those with redshift weights. As shown, the results are consistent with each other in both cases within the uncertainty.

In addition, two BAO papers using the same sample are released as companion papers: \cite{DD18} and \cite{Zhu18}, which are complementary to the isotropic analysis recently presented by \cite{DR14BAO}. These works measure the isotropic and anisotropic BAO in the Fourier and configuration spaces respectively with the optimal redshift weights, and their results are consistent and complementary to each other.

\section{A cosmological implication}
\label{sec:cosmology}

\begin{figure*}
\centering
{\includegraphics[scale=0.35]{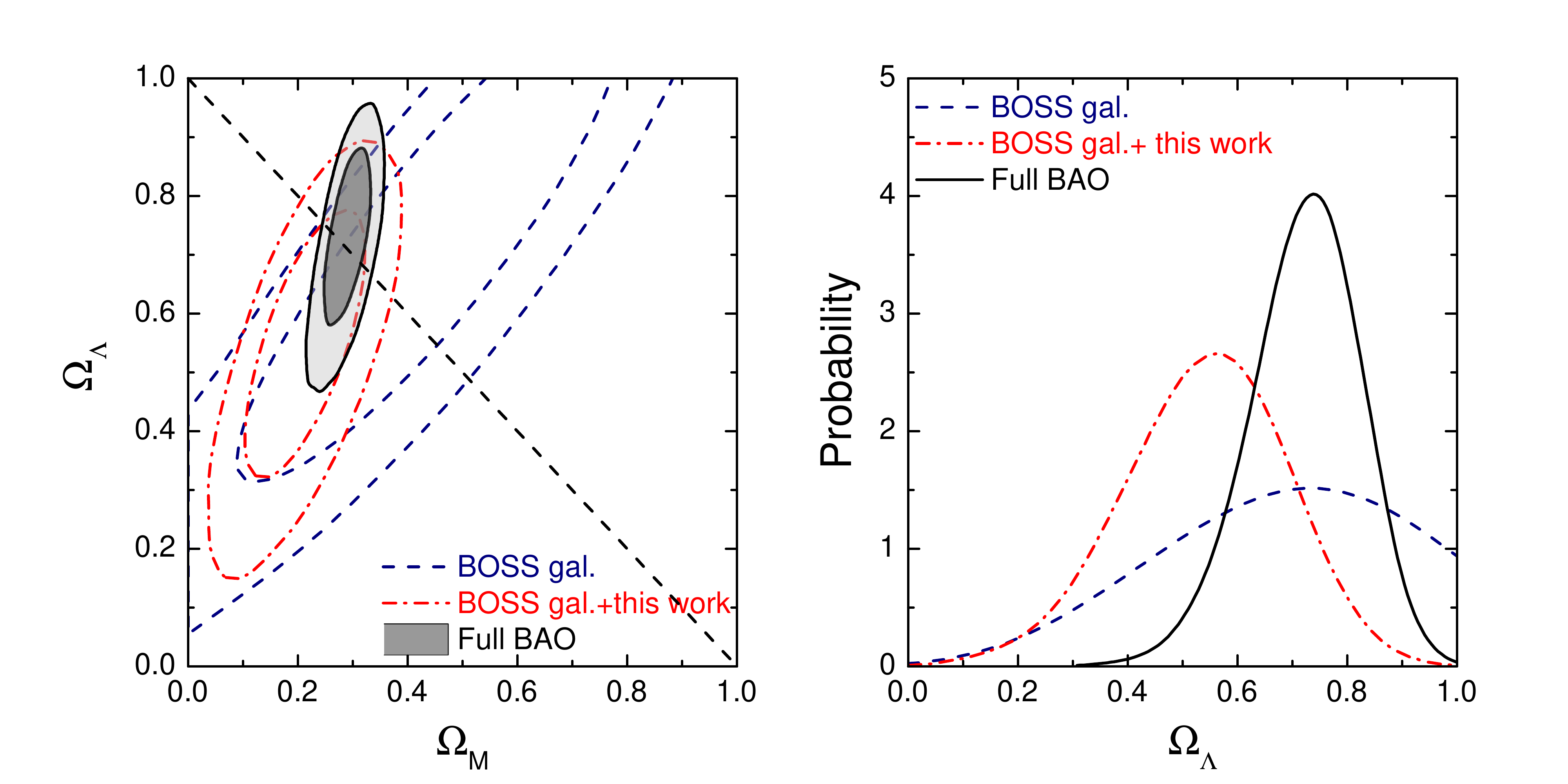}}
\caption{Left: the 68 and 95\% CL contours between $\Omega_{\rm M}$ and $\Omega_\Lambda$ derived from different BAO datasets: BOSS gal. (the consensus BAO measurement in \citet{alam}; BOSS gal. + this work (the BOSS measurement combined with that in this work; Full BAO: BOSS gal. combined with this work and several additional BAO datasets including BOSS DR12 Lyman-$\alpha$ auto- and cross-correlation BAO measurements \citep{dr12lyaf} and the isotropic BAO measurements using MGS \citep{mgs} and 6dFGS \citep{6dF} samples; right: the corresponding one-dimensional posterior distribution.}
\label{fig:Om-OL}
\end{figure*}

\begin{figure}
\centering
{\includegraphics[scale=0.32]{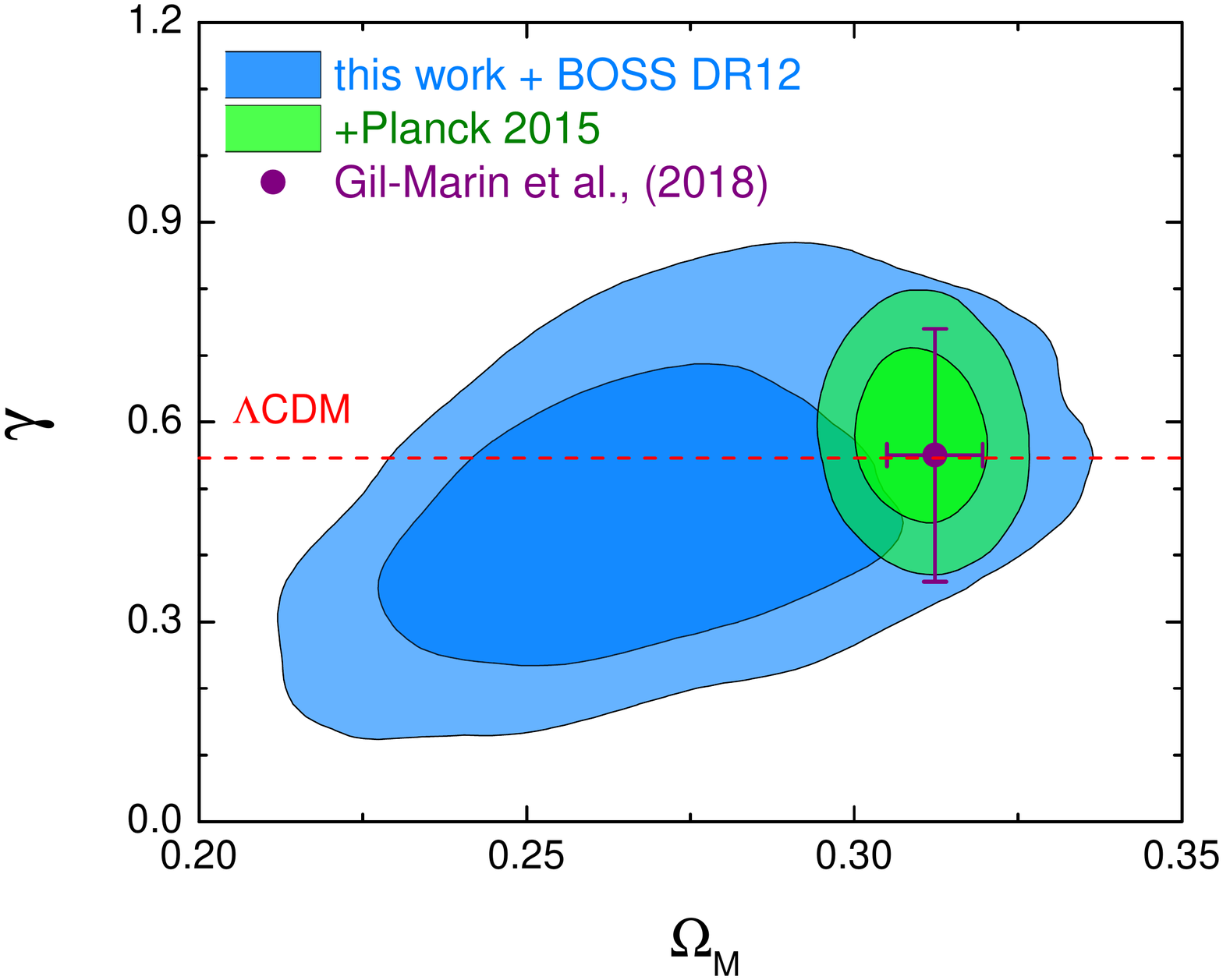}}
\caption{The 68 and 95\% CL contour plots between $\Omega_{\rm M}$ and $\gamma$ derived from this work (presented in Tables \ref{tab:4z} and \ref{tab:corr}) combined with the BOSS DR12 consensus result (blue), and with Planck 2015 combined (green). The data point with error bars shows the measurement from \citet{HGM18}. The horizontal dashed line shows the $\Lambda$CDM prediction, namely, $\gamma=6/11$.}
\label{fig:omm_gamma}
\end{figure}

This subsection is devoted to a cosmological implication of our joint BAO and RSD measurement at four effective redshifts presented in Tables \ref{tab:4z} and \ref{tab:corr}.

We first apply our BAO measurement to calibrate the geometry of the Universe, parametrised by $\Omega_{\rm M}, \Omega_\Lambda$ and $H_0\rd$, using three different BAO data combinations, and present the result in Fig. \ref{fig:Om-OL} and in Table \ref{tab:omol}. 

\begin{table}
\caption{The constraints on $\Omega_{\rm M},\Omega_\Lambda$ and $H_0\rd$ (in unit of ${\rm km/s}$) derived from three BAO data combinations. The signal to noise ratio of $\Omega_\Lambda>0$ (S/N) and the Figure of Merit (FoM) are also shown (the FoM of BOSS gal. is normalised to be unity for the ease of comparison).}
\begin{center}
\begin{tabular}{cccc}
\hline\hline
				&	BOSS gal. &  BOSS gal. + this work & Full BAO \\
\hline
$\Omega_{\rm M}$   & $0.443\pm0.204$	& $0.213\pm0.070$	& $0.289\pm0.028$\\
$\Omega_\Lambda$ & $0.706\pm0.239$	& $0.540\pm0.147$	& $0.722\pm0.098$\\
$H_0\rd$   		& $9.820\pm0.271$	& $9.960\pm0.253$	& $10.166\pm0.206$\\
\hline
S/N				& $2.95$ & $3.67$ & $7.37$ \\
FoM   		        &  $1$	& $3.5$ & $12.1$ \\
\hline\hline
\end{tabular}
\end{center}
\label{tab:omol}
\end{table}%

As shown, our DR14Q BAO measurement combined with DR12 galaxies (BOSS gal. + this work) suggests that dark energy exists at a significance level of $3.67\sigma$, compared to $2.95\sigma$ using BOSS galaxies alone. The Figure of Merit (FoM), which is defined as the square root of the inverse covariance matrix of the $\{\Omega_{\rm M}, \Omega_\Lambda\}$ block, is improved by a factor of $3.5$ by our tomographic DR14Q measurement. Compared to the quasar BAO measurement at a single effective redshift of $1.52$ presented in \cite{DR14BAO}, our measurement is more informative to constrain the geometry of the Universe, namely, the \cite{DR14BAO} measurement improves the BOSS DR12 FoM by a factor of $2$.

We also note that the preferred values of both $\Omega_{\rm M}$ and $\Omega_\Lambda$ derived from this data combination are lower than that favoured by the Planck 2015 measurement by $\sim 1\sigma$, which is confirmed by an independent study in Fourier space in \cite{HGM18} using the same galaxy catalogue. \cite{HGM18} found that, \ba \left\{\Omega_{\rm M}, \Omega_\Lambda\right\} = \left\{0.226^{+0.084}_{-0.093}, 0.55\pm0.14\right\} \ea using BAO measurements in three redshift slices (with effective redshifts of $1.19, 1.50, 1.83$) of the DR14Q  sample. Given the level of uncertainty, we argue that this data combination is still consistent with the Planck observations, and the curvature of the Universe is consistent with zero. However, we will reinvestigate the consistency between quasar BAO and CMB measurements when the final eBOSS quasar survey is completed. 

Combining additional datasets, including the BOSS DR12 Lyman-$\alpha$ auto- and cross-correlation BAO measurements \citep{dr12lyaf} and the isotropic BAO measurements using MGS \citep{mgs} and 6dFGS \citep{6dF} samples, significantly improves the constraint, namely, a Universe without dark energy is excluded at $7.37\sigma$ by these BAO measurements, and the $\{\Omega_{\rm M}, \Omega_\Lambda\}$ constraint using this full BAO dataset is in excellent agreement with the Planck 2015 observations.

We then apply our joint BAO and RSD measurement to constrain the gravitational growth index $\gamma$ together with $\Omega_{\rm M}$, and show the result in Fig. \ref{fig:omm_gamma}. Our measurement combined with BOSS DR12 consensus measurement yields $\gamma=0.469\pm0.148$, which is consistent with the $\Lambda$CDM prediction of $\gamma\sim0.545$. As shown in Table \ref{tab:omol}, this data combination prefers a low $\Omega_{\rm M}$, although the Planck measurement is still within the 68\% CL contour in Fig. \ref{fig:omm_gamma}.

Adding the Planck data tightens the constraint to $\gamma=0.580\pm0.082$, which is consistent with the $\Lambda$CDM prediction. We overplot the 68\% CL uncertainty on $\gamma$ and $\Omega_{\rm M}$ derived from \cite{HGM18} (the ``3z'' result) in Fig. \ref{fig:omm_gamma} for a direct comparison. As shown, our constraint is in general agreement with that in \cite{HGM18}, although our constraint on $\gamma$ is more stringent, probably due to the fact that our RSD measurement is tomographically more informative.  

\section{Conclusion and Discussions}
\label{sec:conclusion}

We present a new and efficient method to extract the lightcone information for both RSD and BAO from galaxy redshift surveys, especially for those covering a wide redshift range.

Based on the optimal redshift weighting scheme, we measure the key parameters for BAO and RSD, namely, $D_{\rm A}, H$ and $f\sigma_8$ for the eBOSS DR14Q sample at four effective redshifts of $z=0.978,1.230,1.526,1.944$, and provide a full data covariance matrix (the key result of this work is presented in Tables \ref{tab:4z} and \ref{tab:corr}). We find an excellent consistency between our measurement and those presented in companion papers, which analyse the same dataset using different methods.

We apply our measurement to constrain the geometry of the Universe, and find that combining our BAO measurement with those from BOSS DR12, MGS and 6dFGS, a Universe without dark energy is excluded at $7.4\sigma$. Our RSD measurement combined with BOSS DR12 and Planck observations yields a constraint of the gravitational growth index, namely, $\gamma=0.580\pm0.082$, which is fully consistent with the GR prediction. 

The method developed in this work can be used to extract the lightcone information from forthcoming deep redshift surveys including DESI \footnote{\url{http://desi.lbl.gov/}}, PFS \footnote{\url{http://pfs.ipmu.jp/}} and Euclid \footnote{\url{https://www.euclid-ec.org/}}, which is crucial for cosmological studies of dark energy \citep{ZhaoDE12, ZhaoDE17}, neutrino mass and modified gravity theories \citep{Zhao:2008bn, MGPCA1}. 
\remove{
References:

DE implications: \cite{ZhaoDE17,ZhaoDE12}
Tomographic analysis: \cite{Zhaotomo16,Wangtomo17,Wangtomo16}

Companion papers:

\cite{RR17, DD17, Zhu17, HGM17, PZ17, Hou17}

eBOSS papers:

\cite{eBOSSZhao16, KD16, MAD15, DR14BAO, DR14, QSObias}

code papers:

\cite{RegPT}
}

\section*{Acknowledgements}

This project has received funding from the European Research Council (ERC) under the European Union's Horizon 2020 research and innovation programme (grant agreement No. 646702 "CosTesGrav"). GBZ, YW and DW are supported by NSFC Grants 11720101004, 11673025 and 11711530207. GBZ is supported by the National Basic Research Program of China (973 Program) (2015CB857004), and by a Royal Society Newton Advanced Fellowship, hosted by University of Portsmouth. YW is supported by a NSFC Grant 11403034, and by a Young Researcher Grant funded by National Astronomical Observatories, Chinese Academy of Sciences. FB is a Royal Society University Research Fellow. G.R. acknowledges support from the National Research Foundation of Korea (NRF) through Grant No. 2017077508 funded by the Korean Ministry of Education, Science and Technology (MoEST), and from the faculty research fund of Sejong University in 2018.

Funding for SDSS-III and SDSS-IV has been provided by
the Alfred P. Sloan Foundation and Participating Institutions.
Additional funding for SDSS-III comes from the
National Science Foundation and the U.S. Department of Energy Office of Science. Further information about
both projects is available at \url{www.sdss.org}.
SDSS is managed by the Astrophysical Research Consortium
for the Participating Institutions in both collaborations.
In SDSS-III these include the University of
Arizona, the Brazilian Participation Group, Brookhaven
National Laboratory, Carnegie Mellon University, University
of Florida, the French Participation Group, the German Participation Group, Harvard University,
the Instituto de Astrofisica de Canarias, the Michigan
State / Notre Dame / JINA Participation Group, Johns
Hopkins University, Lawrence Berkeley National Laboratory,
Max Planck Institute for Astrophysics, Max Planck
Institute for Extraterrestrial Physics, New Mexico State
University, New York University, Ohio State University,
Pennsylvania State University, University of Portsmouth,
Princeton University, the Spanish Participation Group,
University of Tokyo, University of Utah, Vanderbilt University,
University of Virginia, University of Washington,
and Yale University.

The Participating Institutions in SDSS-IV are
Carnegie Mellon University, Colorado University, Boulder,
Harvard-Smithsonian Center for Astrophysics Participation
Group, Johns Hopkins University, Kavli Institute
for the Physics and Mathematics of the Universe
Max-Planck-Institut fuer Astrophysik (MPA Garching),
Max-Planck-Institut fuer Extraterrestrische Physik
(MPE), Max-Planck-Institut fuer Astronomie (MPIA
Heidelberg), National Astronomical Observatories of
China, New Mexico State University, New York University,
The Ohio State University, Penn State University,
Shanghai Astronomical Observatory, United Kingdom
Participation Group, University of Portsmouth, University
of Utah, University of Wisconsin, and Yale University.

This work made use of the facilities and staff of the UK Sciama High Performance Computing cluster supported by the ICG, SEPNet and the University of Portsmouth.
This research used resources of the National Energy Research
Scientific Computing Center, a DOE Office of Science User Facility 
supported by the Office of Science of the U.S. Department of Energy 
under Contract No. DE-AC02-05CH11231.

\newpage

\appendix

\section{The procedure of linearly combining the catalogues}
\label{sec:A}

\begin{figure}
\centering
{\includegraphics[scale=0.35]{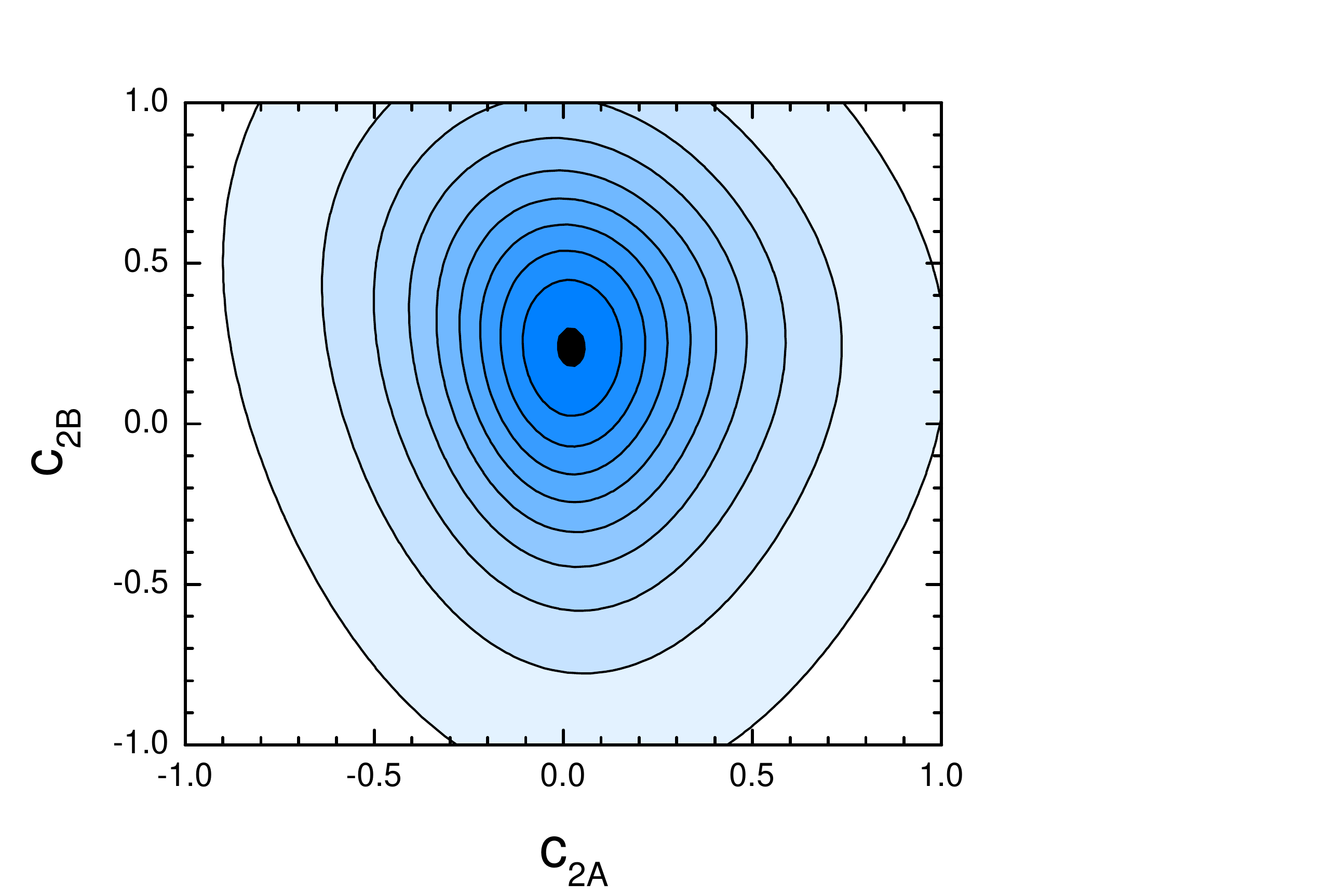}}
\caption{A contour plot of the $c$'s for the redshift-weighted catalogues "2A" (with $\zeff=0.98$) and "2B" (with $\zeff=1.94$), and the colour shows the FoM defined in Eq (\ref{eq:FoM}). The contour lines from inside out illustrate the FoM from maximal to minimal values on linearly uniform intervals, and the black dot in the centre denotes the position where the FoM gets maximised.}
\label{fig:contour_c}
\end{figure}

{\tc

In this section, we provide the procedure of linearly combining the redshift-weighted samples to yield one joint BAO and RSD measurement at a single effective redshift $\zeff=1.526$, in order to compare with the measurement using the unweighted sample at the same effective redshift.
\begin{itemize}
\item Suppose each of the four catalogues is assigned a coefficient $c_i$ ($i$ runs from $1$ to $4$), then the required effective redshift, which is $1.526$ in our case, of the linearly combined sample is,\ba\label{eq:zeffc} \zeff= \left.\left(\sum_{i=1}^4 c_i X_i\right)\right/\left(\sum_{i=1}^4 c_i Y_i\right),\ea and $\Delta_2$ defined in Eq (\ref{eq:d2}) of the combined sample can be calculated as, \ba\label{eq:d2c}  \Delta_2 = \left.\left(\sum_{i=1}^4 c_i Z_i\right)\right/\left(\sum_{i=1}^4 c_i Y_i\right) - z^2_{\rm eff}\ea where \ba \label{eq:XYZ}X_i = \sum_j z_{i,j}w_{i,j}^2; \ Y_i = \sum_j w_{i,j}^2; \ Z_i = \sum_j z^2_{i,j}w^2_{i,j},\ea here the index $j$ runs over all the galaxies in the $i$th catalogue. Evaluate $X,Y,Z$ for each of the four catalogues using Eq (\ref{eq:XYZ});
\item Apply the constraint of \ba\label{eq:cc} \sum_i c_i=1,\ea to properly normalise the linearly combined sample;
\item Compute the FoM of $\{D_A, H, f\sigma_8\}$ of the combined sample as, \ba\label{eq:FoM} {\rm FoM} \equiv \left[{\rm det} \left(C_3\right)\right]^{-1/2}\ea where $C_3$ is the $3\times3$ covariance matrix for $\{D_A, H, f\sigma_8\}$ for the combined sample, which can be derived by linearly combine the sixteen $3\times3$ sub-matrices, denoted as $S$, of $C_{12}$, the full $12\times12$ covariance matrix of the four samples, whose correlation matrix is shown in Figure \ref{fig:corr}. Mathematically, \ba C_3=\sum_{i,j}c_i c_j S_{i,j}.\ea
\item Eqs (\ref{eq:zeffc}) and (\ref{eq:cc}) provide two constraints on the four $\gamma$'s that we are after, then a maximisation of the FoM defined in Eq (\ref{eq:FoM}) while keeping $\Delta_2$ in Eq (\ref{eq:d2c}) negligible can in principle determine the $c$'s.
\end{itemize}

This procedure finds that $c=\{0.02, 0.17, 0.57, 0.24\}$ for the weighted samples with $z_{\rm eff}=0.978, 1.230, 1.526, 1.944$ respectively, and Figure \ref{fig:contour_c} shows a contour plot of the FoM as a function of the two coefficients \footnote{Note that only two of the coefficients are independent given the constraints Eqs (\ref{eq:zeffc}) and (\ref{eq:cc}).}. Due to the correlation among the four catalogues, the trivial solution of $c=\{0, 0, 1, 0\}$ does not maximise the FoM.
} 

\section{The procedure of obtaining the positive redshift weights}
\label{sec:B}

{\tc In practice, we take the following procedures to find the positive-definite redshift weights, shown in the right panels of Figure \ref{fig:weightsvd}, from the original weights, illustrated in the left panels of Figure \ref{fig:weightsvd}.

\begin{itemize}
\item Take the SVD weights $V_1$ and $V_2$ for the monopole;
\item Rotate the V vectors by a linear transformation to obtain new weights W, namely, \ba W_1 &=& V_1\cos\theta - V_2\sin\theta + \lambda,  \nonumber \\
W_2 &=& V_1\sin\theta + V_2\cos\theta + \lambda, \nonumber \\ 
W_3 &=& \lambda. \ea where $\theta$ and $\lambda$ are free parameters to ensure that,

\begin{enumerate}
\item $W_i>0$;
\item The sum of dot-products among the normalised $W_i$'s gets minimised.
\end{enumerate}
\item Repeat this process for the SVD weights for the quadrupole.
\end{itemize}
Note that we use the $2D$ rotation matrix to transform the V vectors, which conserves the orthogonality of V. It is true that the additional shift by $\lambda$ spoils the orthogonality, but this is kept to a minimal level because of the minimisation procedure (ii).
}
\section{The Matlab code for the SVD analysis}

We perform the SVD analysis using the following Matlab code to find the orthogonal redshift weights shown in left panels of Figure \ref{fig:weightsvd}, from the raw redshift weights shown in Figure \ref{fig:weightp0p2}. \\ \\
\noindent\texttt{[m,n]=size(w);}\\ \\ 
\texttt{\% Subtract off the mean of data}\\
\texttt{mn = mean(w,2);}\\
\texttt{w = w-repmat(mn,1,n);}\\ \\ 
\texttt{\% Construct the data matrix X}\\
\texttt{x = w'/sqrt(n-1);}\\ \\ 
\texttt{\% SVD}\\
\texttt{[u,s,pc]=svd(x);}\\ \\
\texttt{\% Perform a projection}\\
\texttt{v = pc'*w;}\\
\texttt{v = v';}\\

\section{The survey window functions}
\label{sec:window}

\begin{figure}
\centering
\includegraphics[scale=0.4]{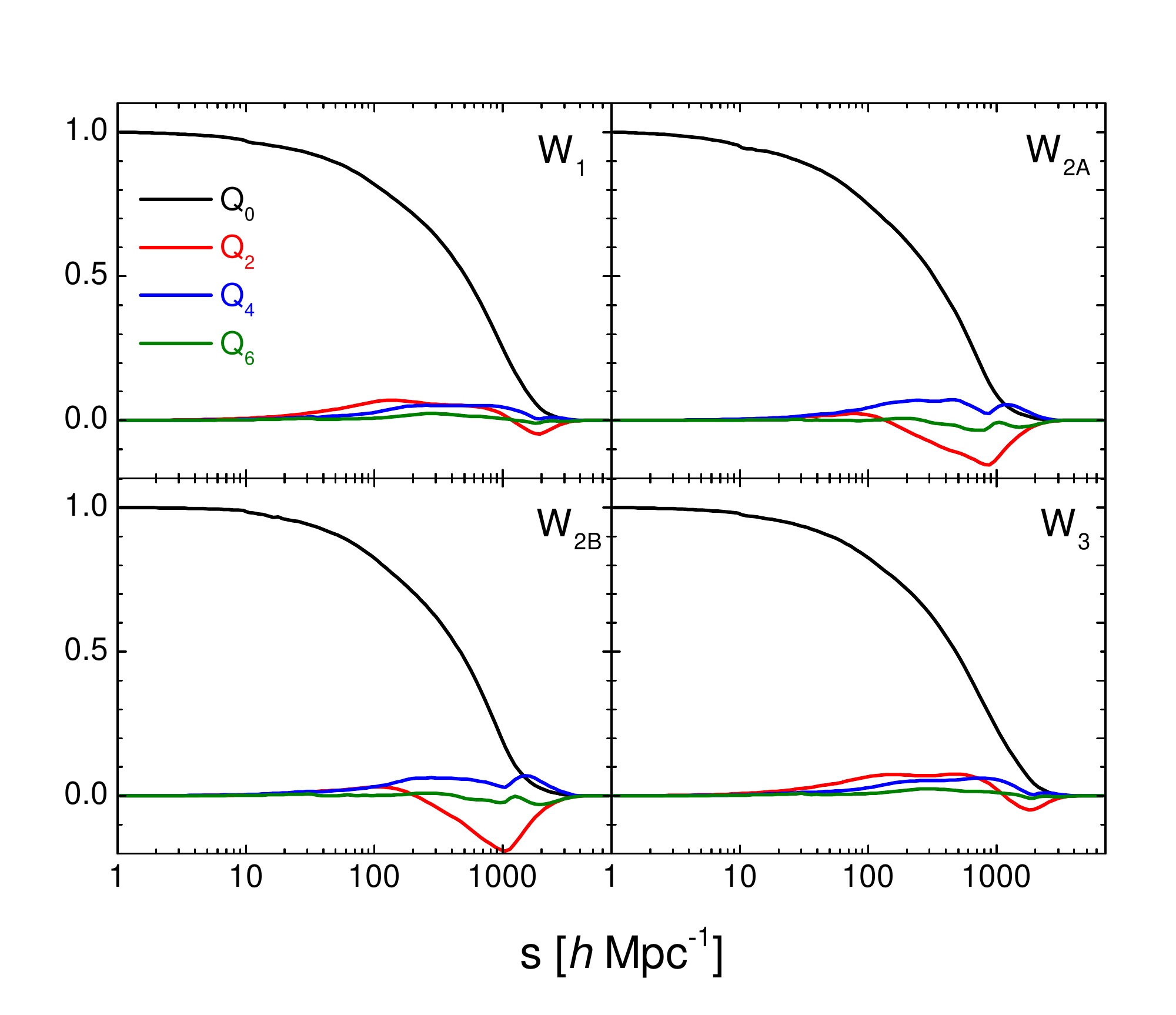}
\caption{The window functions for four redshift-weighted samples, as shown in the legend.}
\label{fig:window}
\end{figure}

The survey window functions for the four redshift-weighted samples are shown in Fig. \ref{fig:window}. These window functions are derived following the method developed in \cite{pkmask}.

\bibliographystyle{mnras}
\bibliography{draft}

\bsp
\label{lastpage}
\end{document}

%% file: tables/widetable1.tex
\begin{table*}
\centering
\caption{The measurement of BAO, RSD and other relevant parameters from the DR14 QSO sample at four effective redshifts. The unit of $D_{\rm M}$ and $D_{\rm V}$ is Mpc, and that of $H$ is km \ s$^{-1}$ \ Mpc$^{-1}$.}
\begin{tabular}{cccccc}
\hline\hline

            &  \multicolumn{4}{c}{DR14 QSO sample}  \\
            &  $z_{\rm eff}=0.978$    & $z_{\rm eff}=1.230$ 				&		$z_{\rm eff}=1.526$  	& $z_{\rm eff}=1.944$ 	\\	
\hline                 
$\alpha_{\bot}$   & $0.939\pm0.169$ &  $1.003\pm0.091$    &   $0.986\pm0.054$  &$1.017\pm0.082$\\
$\alpha_{\|}$       &$1.061\pm0.130$  &  $1.053\pm0.100$   &   $1.095\pm0.094$   &$1.155\pm0.093$\\
$\alpha $            & $0.971\pm0.108$ &$1.017\pm0.056$ &$1.020\pm0.037$ &$1.059\pm0.056$  \\
$\epsilon $            & $0.048\pm0.087$ &$0.018\pm0.054$ &$0.036\pm0.041$ &$0.044\pm0.044$  \\
$D_{\rm A}\times\left(\rfd/\rd\right)$           & $1586.18\pm284.93$ &$1769.08\pm159.67$ &$1768.77\pm96.59$ &$1807.98\pm146.46$  \\
$H\times\left(\rd/\rfd\right)$           & $113.72\pm14.63$ &$131.44\pm12.42$ &$148.11\pm12.75$ &$172.63\pm14.79$  \\
$D_{\rm V}\times\left(\rfd/\rd\right)$           & $2933.59\pm327.71$ &$3522.04\pm192.74$ &$3954.31\pm141.71$ &$4575.17\pm241.61$  \\
$F_{\rm AP}$           & $1.200\pm0.310$ &$1.736\pm0.272$ &$2.212\pm0.265$ &$3.071\pm0.416$  \\
$f\sigma_8$       &$0.379\pm 0.176$  &  $0.385\pm0.099$    &   $0.342\pm0.070$   & $0.364\pm0.106$\\     
$b_1\sigma_8$   &$0.826\pm0.080$  & $0.894\pm0.051$   &   $0.953\pm0.044$     & $1.080\pm0.057$\\   
$b_2\sigma_8$  &$0.460\pm0.684$   & $0.605\pm0.533$    &   $0.704\pm0.507$    & $0.929\pm0.681$\\    
$\sigma_v$        &$3.784\pm1.087$  & $4.732\pm0.761$     &   $5.822\pm0.796$     & $7.591\pm1.127$\\  
\hline
$\chi^2$/DoF    & $56/(58-8)$ & $53/(58-8)$   & $44/(58-8)$ & $40/(58-8)$\\  
$\Delta_2$ & $0.001$ & $0.007$&  $0.141$   &$0.004$ \\  
\hline\hline 

\end{tabular}

%% file: tables/widetable2.tex
\begin{table*}
\centering
\caption{The upper triangular part of the table: the correlation matrix shown in Fig. \ref{fig:corr}; the lower triangular part: the precision matrix. Both the correlation and precision matrices are multiplied by $10^4$ for illustration. The $1/s_i$ column shows {\tc the squareroot of} the reciprocal of the diagonal of the inverse covariance matrix. The dashed lines separate the entries for different effective redshifts for illustration.}
\begin{tabular}{cccccccccccccc}
\hline\hline
      Parameters                                       & \multicolumn{12}{c}{$10^4 \ f_{ij}$ (lower triangular) and $10^4 \ c_{ij}$ (upper) } & $1/s_i$ \\
                                              \hline
$	D_{\rm A}(0.978)	$&$	10000	$&$	3106	$&$	8142	$&$	4656	$&$	1535	$&$	3957	$&$	2662	$&$	920	$&$	2328	$&$	248	$&$	-194	$&$	12	$&$	{\tc 115.6}	$\\
$	H(0.978)	$&$	3399	$&$	10000	$&$	5066	$&$	1543	$&$	5341	$&$	2699	$&$	265	$&$	2934	$&$	1165	$&$	-786	$&$	23	$&$	-377	$&$	{\tc 9.57}	$\\
$	f\sigma_8(0.978)	$&$	-8711	$&$	-5542	$&$	10000	$&$	3740	$&$	2564	$&$	5118	$&$	1718	$&$	1313	$&$	2667	$&$	24	$&$	-192	$&$	-113	$&$	{\tc 0.060}	$\\ \hdashline
$	D_{\rm A}(1.230)	$&$	-5983	$&$	-2613	$&$	6051	$&$	10000	$&$	3993	$&$	8621	$&$	6130	$&$	2421	$&$	5460	$&$	954	$&$	313	$&$	845	$&$	{\tc 32.16}	$\\
$	H(1.230)	$&$	-3509	$&$	-5157	$&$	4773	$&$	5721	$&$	10000	$&$	5994	$&$	1711	$&$	6056	$&$	3409	$&$	-566	$&$	-40	$&$	-300	$&$	{\tc 5.09}	$\\
$	f\sigma_8(1.230)	$&$	5657	$&$	3282	$&$	-6400	$&$	-9462	$&$	-7165	$&$	10000	$&$	4831	$&$	3584	$&$	6288	$&$	510	$&$	226	$&$	722	$&$	{\tc 0.016}	$\\ \hdashline
$	D_{\rm A}(1.526)	$&$	2862	$&$	1198	$&$	-3048	$&$	-8204	$&$	-4824	$&$	8032	$&$	10000	$&$	3888	$&$	8574	$&$	4257	$&$	1941	$&$	3919	$&$	{\tc 20.13}	$\\
$	H(1.526)	$&$	1946	$&$	1226	$&$	-2276	$&$	-5180	$&$	-7167	$&$	6212	$&$	6111	$&$	10000	$&$	6015	$&$	1150	$&$	3897	$&$	2156	$&$	{\tc 5.21}	$\\
$	f\sigma_8(1.526)	$&$	-3002	$&$	-1304	$&$	3253	$&$	7898	$&$	5798	$&$	-8422	$&$	-9419	$&$	-7516	$&$	10000	$&$	3519	$&$	2623	$&$	4721	$&$	{\tc 0.011}	$\\ \hdashline
$	D_{\rm A}(1.944)	$&$	-1454	$&$	-18	$&$	1360	$&$	5098	$&$	3331	$&$	-5176	$&$	-6910	$&$	-4718	$&$	6758	$&$	10000	$&$	3306	$&$	8127	$&$	{\tc 58.0}	$\\
$	H(1.944)	$&$	-1041	$&$	-690	$&$	1255	$&$	3434	$&$	4368	$&$	-4012	$&$	-4382	$&$	-6340	$&$	5166	$&$	4855	$&$	10000	$&$	5687	$&$	{\tc 8.99}	$\\
$	f\sigma_8(1.944)	$&$	1790	$&$	363	$&$	-1758	$&$	-5487	$&$	-4130	$&$	5933	$&$	7069	$&$	5987	$&$	-7707	$&$	-8799	$&$	-6752	$&$	10000	$&$	{\tc 0.032}	$\\\hline\hline
\end{tabular}

%% file: tables/widetable3.tex
\begin{table*}
\centering
\caption{The measurement of BAO and RSD parameters at the effective redshift $1.526$ with the redshift weights. The notations are the same as those in Table \ref{tab:corr}.}
\begin{tabular}{ccccccc}
\hline\hline
   Parameters    &      Mean  &  Uncertainty &    \multicolumn{3}{c}{$10^4 \ f_{ij}$ (lower triangular) and $10^4 \ c_{ij}$ (upper) } & $1/s_i$ \\
 \hline
$D_{\rm A}(1.526)$    & $1774.59$ & $94.83$ & $10000$& $3619$ & $8356$ &   $50.33$                                   \\
$	H(1.526)	$        & $150.32$ & $10.50$ & $2572$ & $10000$ & $5718$ &    $8.32$                                  \\
$f\sigma_8 \ (1.526)$  & $0.356$ & $0.067$ & $-8220$ & $-5261$ & $10000$ &   $0.031$                                   \\
\hline\hline
\end{tabular}